\documentclass[letter]{aa}  

\usepackage{graphicx}
\usepackage{txfonts}

\begin{document}

\title{A major asymmetric ice trap in a planet-forming disk: \\ I. Formaldehyde and methanol}

\author{Nienke van der Marel 
\inst{1,2}
\and
Alice S. Booth \inst{3}
\and
Margot Leemker \inst{3}
\and
Ewine F. van Dishoeck \inst{3,4}
\and
Satoshi Ohashi \inst{5}
}

\institute{Physics \& Astronomy Department, 
University of Victoria, 
3800 Finnerty Road, 
Victoria, BC, V8P 5C2, 
Canada
\and
Banting Research fellow
\and
Leiden Observatory, Leiden University, 2300 RA Leiden, the Netherlands
\and
Max-Plank-Institut fur Extraterrestrische Physik, Giessenbachstraße 1, D-85748 Garching, Germany
\and
RIKEN Cluster for Pioneering Research, 2-1, Hirosawa, Wako-shi, Saitama 351-0198, Japan}

\date{Received April, 2021; accepted June, 2021}

\abstract
{The chemistry of planet-forming disks sets the exoplanet atmosphere composition and the prebiotic molecular content. Dust traps are of particular importance as pebble growth and transport are crucial for setting the chemistry where giant planets are forming.} 
{The asymmetric Oph~IRS~48 dust trap located at 60 au radius provides a unique laboratory for studying chemistry in pebble-concentrated environments in warm Herbig disks with low gas-to-dust ratios down to 0.01.} 
{We use deep ALMA Band~7 line observations to search the IRS~48 disk for H$_2$CO and CH$_3$OH line emission, the first steps of complex organic chemistry.} 
{We report the detection of 7 H$_2$CO and 6 CH$_3$OH lines with energy levels between 17 and 260 K. The line emission shows a crescent morphology, similar to the dust continuum, suggesting that the icy pebbles play an important role in the delivery of these molecules. Rotational diagrams and line ratios indicate that both molecules originate from warm molecular regions in the disk with temperatures $>$100 K and column densities $\sim10^{14}$ cm$^{-2}$ or a fractional abundance of $\sim10^{-8}$ and with H$_2$CO/CH$_3$OH$\sim$0.2, indicative of ice chemistry. Based on arguments from a physical-chemical model with low gas-to-dust ratios, we propose a scenario where the dust trap provides a huge icy grain reservoir in the disk midplane or an `ice trap', which can result in high gas-phase abundances of warm COMs through efficient vertical mixing.} 
{This is the first time that complex organic molecules have been clearly linked to the presence of a dust trap. These results demonstrate the importance of including dust evolution and vertical transport in chemical disk models, as icy dust concentrations provide important reservoirs for complex organic chemistry in disks.}


\maketitle

\section{Introduction} \label{sec:intro}
Protoplanetary disks around young stars are the birth cradles of planets, and the chemical composition in these disks sets the exoplanet atmospheric composition and the formation of prebiotic molecules on their surfaces \citep{Ehrenfreund2000, Oberg2020}. So far, mostly simple molecules have been detected in disks \citep[e.g.][]{Dutrey1997,Thi2004,Oberg2010,Oberg2015,Walsh2016}, and their abundances are set by photodissociation in the surface layers and freeze-out in the midplane \citep{Bergin2007}. Complex organic molecules (COMs) may be present but are expected to be mostly locked up in ices. CO ice chemistry is crucial for the formation of complex organic molecules which can be thermally released into the gas-phase \citep{Herbst2009}. For Herbig disks, COMs cannot form in situ since they are warm and lack a large CO-ice reservoir \citep{Agundez2018}. Surprisingly, CH$_3$OH was recently detected in the Herbig disk HD100546 \citep{Booth2021}. This detection can be understood when CH$_3$OH ice is inherited from earlier stages, followed by radial transport and sublimation at its iceline. Pebble growth and transport are known to play an important role in the chemical composition of disks and resulting exoplanet atmospheres \citep{Cridland2017,Krijt2020}.  The connection between pebbles and ice chemistry can be studied directly in so-called dust traps (concentrations of dust grains), possibly revealing a much richer chemistry as dust rings are colder in the midplane \citep{Alarcon2020} while exposed dust cavity walls can reveal sublimated midplane products  \citep{Cleeves2011,Mulders2011}.

Dust traps are thought to be the main explanation for the observed narrow dust rings and asymmetries in high-resolution ALMA observations revealing also a segregation between gas and dust \citep[e.g.][]{vanderMarel2013,Perez2014,Andrews2018}. Pressure bumps at gap edges trap larger dust grains due to drag forces between gas and dust \citep{Weidenschilling1977}, which can explain the appearance of dust rings in protoplanetary disks as the dust is prevented from drifting inwards \citep{Pinilla2012b}. In some cases, the pressure bump can become susceptible to the Rossby Wave Instability and form long-lived vortices \citep{BargeSommeria1995}, which trap the dust in the azimuthal direction. The Herbig disk Oph~IRS~48 is a textbook example of such a dust trap showing an asymmetric dust concentration south of the star \citep{vanderMarel2013,vanderMarel2015vla}, and thus an ideal target for studying complex organic chemistry in a pebble-concentrated environment.

\begin{table*}[!ht]
    \centering
    \caption{Detected molecular lines, line properties and disk integrated fluxes}
    \label{tab:lines}
    \begin{tabular}{lllrllr}
    \hline
    Molecule&Transition&Rest frequency&$E_u$&$g_u$&$\log A_{ul}$&$F_{int}$\\
    &&(GHz)&(K)&&&(mJy km s$^{-1}$)\\
    \hline
o-H$_2$CO	&	5$_{1,5}$-4$_{1,4}$	&	351.768648	&	62	&	33	&	-2.92013	&	836	\\
o-H$_2$CO	&	5$_{3,3}$-4$_{3,2}$	&	364.275141	&	158	&	33	&	-3.05097	&	391	\\
o-H$_2$CO	&	5$_{3,2}$-4$_{3,1}$	&	364.288914	&	158	&	33	&	-3.05065	&	574	\\
p-H$_2$CO	&	5$_{0,5}$-4$_{0,4}$	&	362.736024	&	52	&	11	&	-2.86264	&	577	\\
p-H$_2$CO	&	5$_{2,4}$-4$_{2,3}$	&	363.945876	&	100	&	11	&	-2.93377	&	377	\\
p-H$_2$CO	&	5$_{4,2}$-4$_{4,1}$/5$_{4,1}$-4$_{4,0}$$^a$	&	364.103257	&	241	&	11	&	-3.30139	&	99	\\
\hline
a-CH$_3$OH	&	14$_{1,13}$-14$_{0,14}$	&	349.106997	&	260	&	116	&	-3.35603	&	198	\\
a-CH$_3$OH	&	1$_{1,1}$-0$_{0,0}$ 	&	350.905100	&	17	&	12	&	-3.47949	&	89	\\
e-CH$_3$OH	&	4$_{0,4}$-3$_{-1,3}$	&	350.687651	&	36	&	36	&	-4.06195	&	215	\\
e-CH$_3$OH	&	7$_{-2,6}$-6$_{-1,5}$	&	363.739868	&	87	&	60	&	-3.76767	&	141	\\
e-CH$_3$OH	&	8$_{1,7}$-7$_{2,5}$ 	&	361.852195	&	105	&	68	&	-4.11248	&	125	\\
e-CH$_3$OH	&	11$_{0,11}$-10$_{1,9}$	&	360.848946	&	166	&	92	&	-3.91831	&	155	\\
\hline
\end{tabular}
\\
$^a$ Lines are blended.\\
Rest frequencies and other properties are taken from CDMS: $E_u$ is the upper energy level, \\ 
$g_u$ the degeneracy and $A_{ul}$ the Einstein~$A$-coefficient. 
\end{table*}

The presence of warm H$_2$CO in the IRS~48 disk was discovered by \citet{vanderMarel2014}. H$_2$CO is a precursor of more complex organic molecules such as CH$_3$OH through CO ice hydrogenation \citep[e.g.][]{Watanabe2002,Fuchs2009}. The morphology appeared to be cospatial with the dust crescent, but the detection was tentative. CH$_3$OH was not detected in this work, but upper limits were derived. The H$_2$CO/CH$_3$OH abundance ratio can be used as tracer for the formation mechanism \citep{Garrod2006}, since CH$_3$OH can only be formed efficiently through ice chemistry, whereas H$_2$CO has both an ice- and gas-phase route \citep[e.g.][]{Walsh2014}. Gas-phase dominated H$_2$CO formation implies a ratio$>$1 and ice-phase a ratio$<$1. However, the derived  H$_2$CO/CH$_3$OH ratio from \citet{vanderMarel2014} of $\gtrsim$0.3 was inconclusive about the formation mechanism.

Whereas H$_2$CO has been routinely detected in a range of protoplanetary disks \citep[e.g.][]{Pegues2020}, CH$_3$OH has only been detected in the TW~Hya disk \citep{Walsh2016}, the young IRAS04302 disk \citep{Podio2020} and outburst disk V883~Ori \citep{vantHoff2018,Lee2019} and recently in the Herbig disk HD100546 \citep{Booth2021}. 

In this work, we present the detection of multiple CH$_3$OH and H$_2$CO transitions in the IRS~48 system, only the second Herbig disk with observed COMs and the first disk where the COM production can be linked directly to the dust trap. 

\section{Observations}
Oph~IRS~48 is an A0~star located in the Ophiuchus cloud at a distance of 135~pc \citep{Gaia2018}. This disk, inclined at 50$^{\circ}$, shows an asymmetric millimeter-dust concentration at 60~au, in contrast with a full ring in gas and small dust grains \citep{vanderMarel2013} and has an estimated gas disk mass of only 0.6 $M_{\rm Jup}$ \citep{vanderMarel2016-isot}. IRS~48 has been observed using the Atacama Large Millimeter/submillimeter Array (ALMA) in Band~7 in polarization mode in Cycle~5 in August 2018 (2017.1.00834.S, PI:Adriana Pohl). The continuum polarization data are presented by \citet{Ohashi2020}, and the main calibration and reduction process is described in detail in that work. The total on-source integration time is 89 minutes. The continuum is subtracted in the uv-plane using the CASA task \texttt{uvcontsub} with a first order polynomial. The spectral setup contains four spectral windows at 349.7, 351.5, 361.6 and 363.5~GHz, with a bandwidth of 1875~GHz in each window and a channel width of 1953~kHz or $\sim$1.6~km~s$^{-1}$. 

Seven H$_2$CO and six CH$_3$OH transitions were identified using the matched-filter-technique \citep{Loomis2018}, listed in Table \ref{tab:lines}, for $E_u$ levels between 17 and 260~K. H$_2$CO transitions are identified as ortho (o-) and para (p-) transitions, respectively. Each line was imaged using the \texttt{tclean} task at the channel resolution, using natural weighting. The final channel cubes have a beam size of 0.63$\times$0.50" and a rms noise of $\sigma_{\rm rms}\sim$1.2~mJy~beam$^{-1}$~channel$^{-1}$. The brightest lines (H$_2$CO~5$_{0,5}-4_{0,4}$ and CH$_3$OH~4$_{0,4}-3_{1,3}$) are also imaged using Briggs weighting with a robust of 0.5 for a resolution of 0.55$\times$0.44". 

Spectra are extracted from the naturally weighted cubes using Keplerian masking (using $d$=135 pc, $i$=50$^{\circ}$ and $M_*$=2.0 $M_{\odot}$) and presented in Figure \ref{fig:spectra}. All cubes show a clear Keplerian pattern along the southern part of the disk. Some lines are located adjacent to other lines, i.e. the H$_2$CO~5$_{3,3}$-4$_{3,2}$ and 5$_{3,2}$-4$_{3,1}$ transitions, so their line wings overlap in 2 channels. 

The disk-integrated spectra are resolved even at our low spectral resolution, ranging from -2 to 12~km~s$^{-1}$ with $v_{\rm source}$=4.55~km~s$^{-1}$. The CH$_3$OH lines appear to have somewhat more prominent line wings (corresponding to an inner 30~au radius) than the H$_2$CO lines. The spectra are integrated over the entire individual profiles (avoiding overlap with adjacent features) and the disk integrated fluxes are reported in Table \ref{tab:lines}. The H$_2$CO~5$_{4,2}$-4$_{4,1}$ and 5$_{4,1}$-4$_{4,0}$ fluxes are computed by dividing their shared flux by 2. The integrated fluxes are detected with a range between 5 and 42$\sigma_{\rm int}$, with $\sigma_{\rm int}\sim$20~mJy~km~s$^{-1}$, whereas the calibration uncertainty is 10\%. 

\begin{figure*}[!ht]
    \centering
    \includegraphics[width=\textwidth]{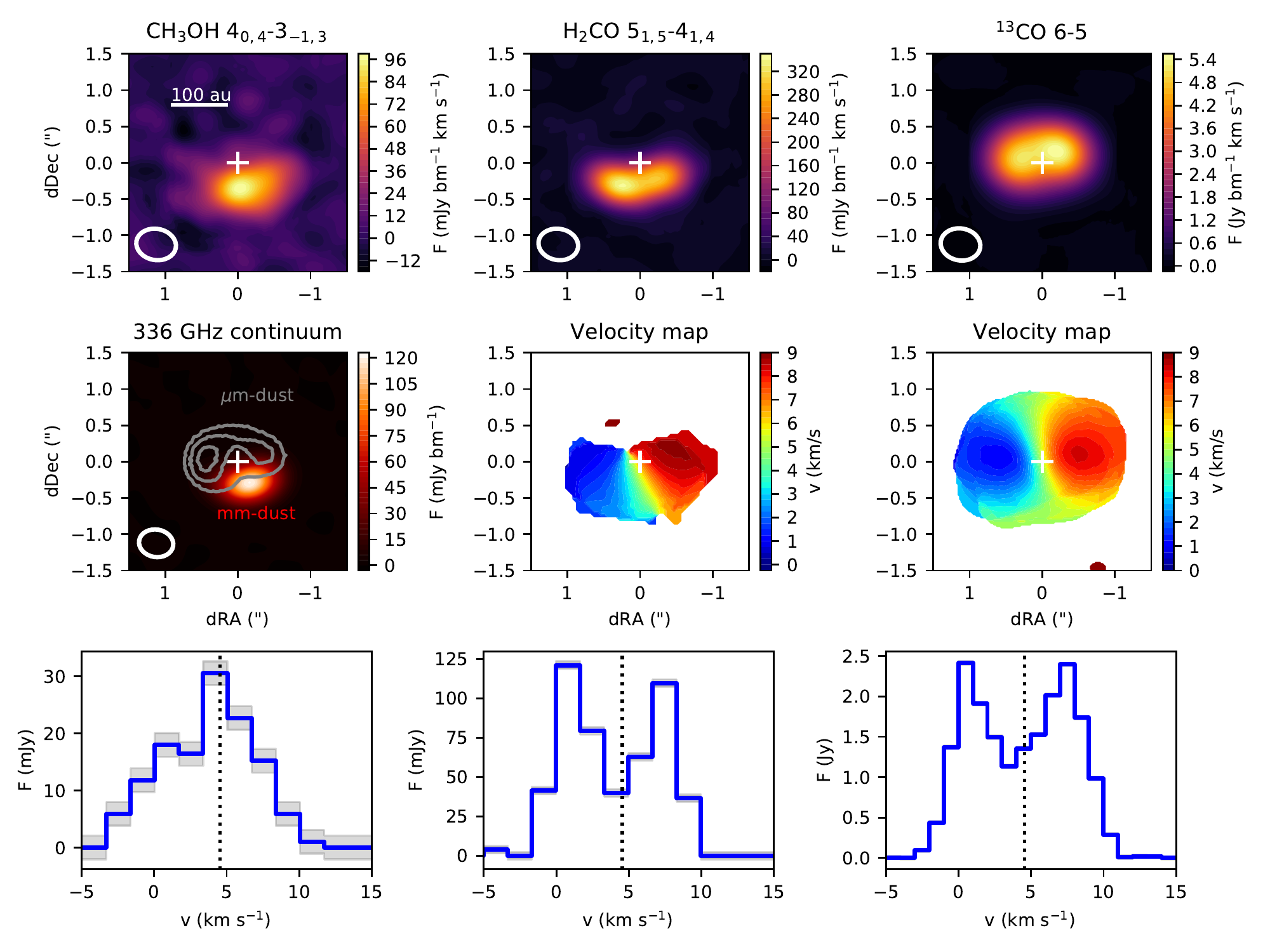}
    \caption{Overview of the brightest H$_2$CO and CH$_3$OH lines using Briggs weighting, the 336~GHz continuum and the $^{13}$CO~6-5 map with a similar beam size for comparison. The top row shows the zero-moment maps, the middle row the continuum and the first moment maps and the bottom row the disk integrated spectra. The $\mu$m-sized grain distribution as traced by 19 $\mu$m VISIR data \citep{Geers2007} is indicated in the continuum image. The source velocity is indicated by a vertical dotted line. The grey shades indicate the noise levels in the spectra.}
    \label{fig:mommaps_overview}
\end{figure*}

Zero-moment maps are created using Keplerian masking and presented in Figure \ref{fig:mommaps}. The central position is set at J2000~16h27m37.180s, -24$^{\circ}$30'35.48", based on Gaia~DR2 \citep{Gaia2018}. Spectra and moment maps of the two brightest lines are presented in Figure \ref{fig:mommaps_overview}.

\section{Results}
Both H$_2$CO and CH$_3$OH lines are firmly detected. IRS~48 is the second known Herbig disk with a detection of CH$_3$OH, following HD~100546 \citep{Booth2021}.  It is immediately clear that both molecules follow the dust trap morphology (Figure \ref{fig:mommaps_overview}), in contrast with $^{13}$CO which shows a full disk ring, just like the small grains \citep{vanderMarel2013}. This confirms the findings of \citet{vanderMarel2014} of their suggested location of the H$_2$CO emission. 

Figure \ref{fig:mommaps_overview} presents the data for the two brightest line transitions: the H$_2$CO~$5_{1,5}-4_{1,4}$ and the CH$_3$OH~$4_{0,4}-3_{0,3}$ lines. The maps are compared with the 355~GHz continuum from the same dataset and with the $^{13}$CO 6--5 intensity maps. The $^{13}$CO~data are taken from \citet{vanderMarel2016-isot} and imaged using uv-tapering for a similar beam size as the Band~7 data presented here. The first-moment maps of the H$_2$CO and CH$_3$OH emission are consistent with Keplerian motion along the southern half of the disk. 

\begin{figure}[!ht]
    \centering
    \includegraphics[width=0.5\textwidth]{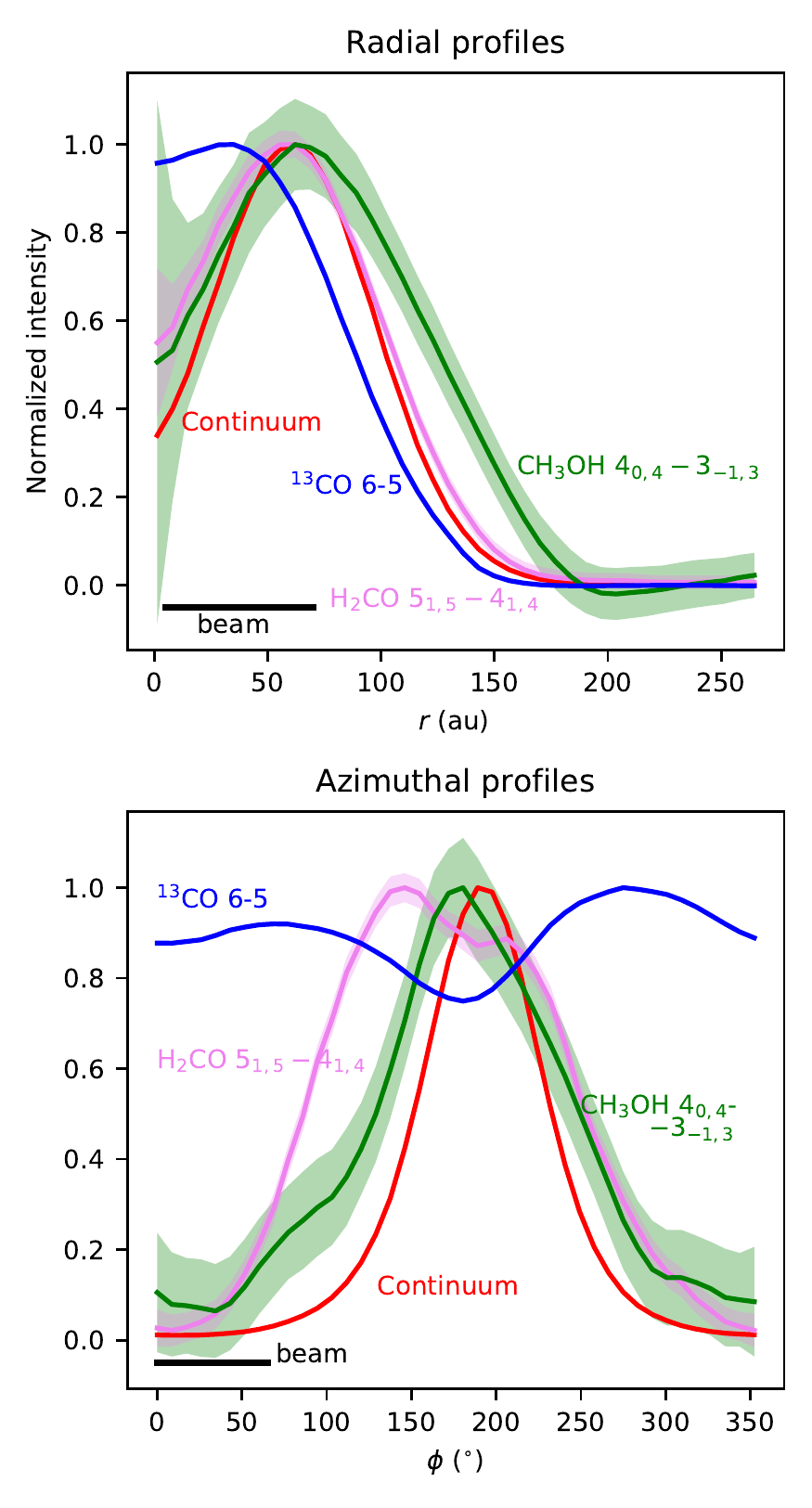}
    \caption{Radially and azimuthally averaged profiles of the brightest H$_2$CO and CH$_3$OH lines from the Briggs weighted images, combined with the profiles from $^{13}$CO 6-5 and the 355 GHz continuum imaged at the same beam size of 0.55$\times$0.44".}
    \label{fig:curves}
\end{figure}

A comparison between the images in both the radial and azimuthal directions is presented in Figure \ref{fig:curves}. The profiles are extracted by deprojecting the zero-moment maps assuming a position angle of 100$^{\circ}$ and an inclination of 50$^{\circ}$ \citep{Bruderer2014}. The azimuthal profile is extracted at the dust peak radius of 62~au with a radial width of 60 au and the radial profile at the peak $\phi$ of 192$^{\circ}$ East-of-North with an azimuthal width of 100$^{\circ}$. For $^{13}$CO, the data are extracted around the peak radius of 35~au and the peak $\phi$ of 269$^{\circ}$. 

The azimuthal profiles show that H$_2$CO is azimuthally more extended than CH$_3$OH and is trailing the dust trap, whereas CH$_3$OH is similar in width as the continuum emission. In contrast, the $^{13}$CO emission is present along the entire ring, with a dip around the continuum peak due to continuum oversubtraction (the Band~9 continuum peak is azimuthally shifted w.r.t. to the Band~7 continuum due to different grain sizes traced). The Band~7 line data presented here are only moderately affected by continuum oversubtraction. Radially, the H$_2$CO profile is cospatial with the continuum emission, although the emission remains radially unresolved. In contrast, the CH$_3$OH profile appears somewhat further extended outwards and based on the line wings, also inwards. The $^{13}$CO emission peaks radially inside the dust continuum peak. 

\section{Analysis}
With multiple line transitions it is possible to derive the column density and excitation temperature for both molecules under the assumption of LTE and optically thin emission (or a correction for optical depth). The optical depth is determined first using the expected emission from the emitting area. As the zero-moment maps of H$_2$CO and CH$_3$OH are marginally resolved, the emitting area cannot be reliably determined from these images. Instead, the emitting area is determined using the high resolution (0.18$\times$0.14") Band~7 continuum image \citep[][and Figure \ref{fig:highres}]{Francis2020}, with the underlying assumption that the H$_2$CO and CH$_3$OH emission follow the morphology of the dust crescent. Although the H$_2$CO emission is azimuthally more extended than the continuum, the difference is marginal ($\sim$20\% in the convolved images) and can be ignored. The emitting area of the high-resolution continuum is 1.4$\cdot10^{-11}$ sr with a 5$\sigma$ threshold.

\begin{figure}[!ht]
    \centering
    \includegraphics[width=0.5\textwidth]{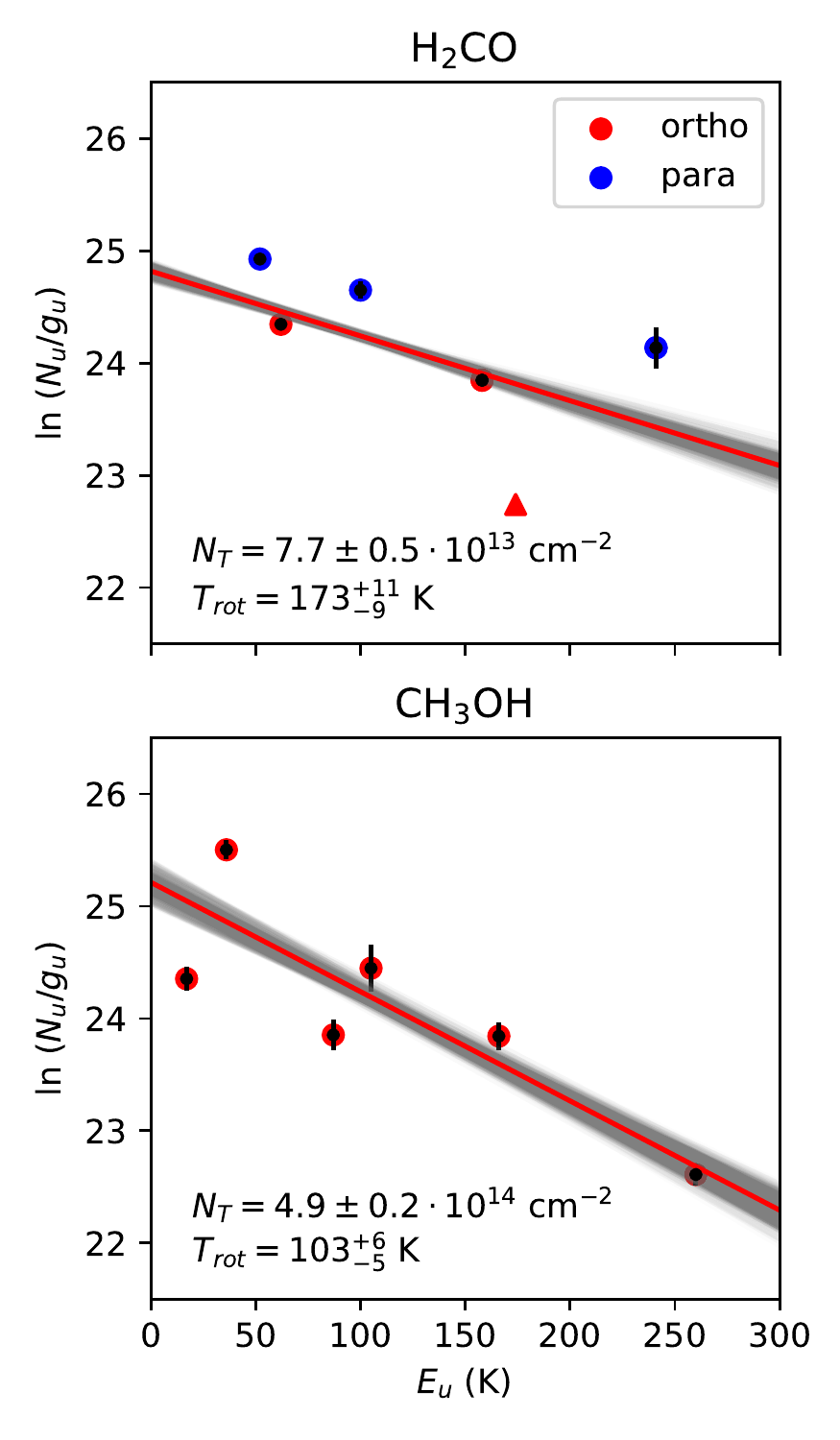}
    \caption{Rotational diagrams of H$_2$CO and CH$_3$OH using the integrated fluxes from this study, assuming optically thin emission. The integrated flux of the H$_2$CO~9$_{1,8}$-8$_{1,7}$ transition from \citet{vanderMarel2014} is included as a lower limit. The red line provides the best fit through the data points and the grey lines are draws from the posterior distribution from the fitting (Figure \ref{fig:posteriors}). }
    \label{fig:rotdiagram}
\end{figure}

Using this emitting area, the optical depth $\tau$ and the column densities of individual levels $N_u$ are estimated following the excitation equations in \citet{Loomis2018}. All lines are optically thin with $\tau<0.1$ for $T>$100 K. The ortho-to-para ratio of H$_2$CO in the degeneracies and partition functions is taken as 3, and the A/E ratio of CH$_3$OH as 1. The column density $N_T$ and temperature $T_{\rm rot}$ are estimated by fitting the rotational diagrams using the \texttt{emcee} package to compute the posterior distributions \citep{DFM2013}, assuming optically thin emission. Our best-fit results are shown in Figure \ref{fig:rotdiagram} and indicate an average column density of 7.7$\pm0.5\cdot10^{13}$ cm$^{-2}$ and 4.9$\pm0.2\cdot10^{14}$~cm$^{-2}$, and a rotational temperature of 173$^{+11}_{-9}$ and 103$^{+6}_{-5}$~K for H$_2$CO and CH$_3$OH, respectively (see Figure \ref{fig:posteriors}). This implies that the temperature of H$_2$CO is higher than CH$_3$OH and the abundance ratio H$_2$CO/CH$_3$OH is 0.16$\pm0.01$, much lower than found for other disks \citep{Booth2021}, and suggestive of ice-dominated chemistry. However, it is possible that the transitions trace multiple regimes with different temperatures in the disk. Furthermore, the ortho-to-para ratio obtained by fitting the ortho and para lines separately is $<$3, which is potentially pointing to an ice formation route as well \citep{TwS2021}, but could also be caused by optical depth (see Appendix \ref{sec:mcmc}). The gas surface density derived by \citet{vanderMarel2016-isot} at 60 au radius corresponds to $N_{\rm H_2}\sim1.6\cdot10^{22}$~cm$^{-2}$, so the relative abundances of H$_2$CO and CH$_3$OH are $\sim10^{-8}$ w.r.t. H$_2$, consistent with previous estimates by \citet{vanderMarel2014}. 

The excitation temperature (rotational temperature) is equal to the kinetic temperature under the assumption of LTE at high densities. H$_2$CO lines are particularly good diagnostics of kinetic temperature, since radiative transitions are not allowed between the different $K_p$ ladders. The relative populations of those ladders are therefore dominated by collisions only \citep{Mangum1993,vanDishoeck1993}. The LTE assumption and corresponding kinetic temperature can be tested using a calculation of the balance between excitation and de-excitation using RADEX \citep{vanderTak2007}. Collisional rate coefficients are taken from the LAMDA database for the individual molecules \citep{Rabli2010,Wiesenfeld2013} as summarized in \citet{Schoier2005}. We compute line ratios for a range of temperatures and H$_2$ densities and a molecular column density of 10$^{14}$~cm$^{-2}$ shown in Figure \ref{fig:radex} following \citet{vanDishoeck1995}.

The H$_2$ densities in the midplane and molecular layers of IRS~48 are $\sim10^{6-8}$~cm$^{-3}$ \citep{Bruderer2014}. In this regime, the line ratios are only sensitive to temperature, with typical inferred values of 200$\pm$50~K for H$_2$CO and 100$\pm$20~K for CH$_3$OH, confirming that the H$_2$CO emission originates from a warmer layer (Fig.~\ref{fig:radex}). 

\section{Discussion and conclusions}
The strong detection of CH$_3$OH and its precursor H$_2$CO in IRS~48 challenge current chemical disk models, which predict that Herbig disks cannot form COMs in situ due to their warmer midplane \citep{Agundez2018}. IRS~48 is the second mature Herbig disk with a CH$_3$OH detection, following HD100546 \citep{Booth2021}. The derived H$_2$CO/CH$_3$OH abundance ratio of $\sim$0.2 in IRS~48 indicates that ice chemistry must be the primary formation mechanism. The warm excitation temperatures $>$100~K indicates that the emission does not originate from the disk midplane which has a temperature of $\sim$70~K at 60~au \citep{Bruderer2014}. The continuum brightness temperature at 355~GHz is 27~K, providing a lower limit. H$_2$CO may originate from slightly higher layers than CH$_3$OH considering its rotational temperature.

An important clue for the origin of the COM chemistry in IRS~48 is the striking crescent morphology of the emission, which is following the shape of the dust continuum. The asymmetric dust continuum has been interpreted as a dust trap, based on the comparison between large grains, small grains and gas \citep{vanderMarel2013}. The high degree of chemical complexity may thus be related to special physical conditions there. Large grains concentrate in a dust trap and grow efficiently to larger sizes due to the higher dust concentration and lower destructive collision efficiency  \citep{Weidenschilling1977,Brauer2008,Pinilla2012a}. Small grains are still continuously produced by fragmentation. The dust trap thus provides a large reservoir of icy dust grains, and if these have been either radially transported from the outer part of the disk or inherited from the early, colder stages they might be rich in ices \citep{Krijt2020,Booth2021}. The dust trap thus acts as an `ice trap' of large icy dust grains, as previously suggested for TW Hya \citep{Walsh2016}. 
Considering typical interstellar ice abundances of CH$_3$OH of $3\cdot10^{-6}$ \citep{Boogert2015}  and model COM ice abundances in the disk midplane of $\geq$10$^{-6}$ \citep{Walsh2014}, only a fraction of the ice content is sublimated.

The dust density distribution in the dust trap plays a crucial role here, as dust grains limit the UV field penetration in the disk, lowering the dust and gas temperature \citep{Bruderer2012}. \citet{Ohashi2020} derived a dust surface density as high as 2-8~g~cm$^{-2}$ at the dust trap radius based on polarization continuum measurements and constraints from the centimeter emission in IRS~48. This is well above the  gas surface density of 0.07~g~cm$^{-2}$ derived from CO isotopologues and DALI modelling by \citet{vanderMarel2016-isot}, implying a dust-to-gas~ratio $\gg$1. Using a series of physical-chemical models with different dust surface densities, we estimate the temperature in the disk at the location of the dust trap. Two sets of models are run: first, only the dust density of the large grains in the midplane is increased (settled models) and second, the dust density is increased throughout the column (full models). The details of the vertical and radial structure of gas and dust of the models are described in Appendix \ref{sct:temperature}.

\begin{figure*}[!ht]
\centering
\includegraphics[width=\textwidth]{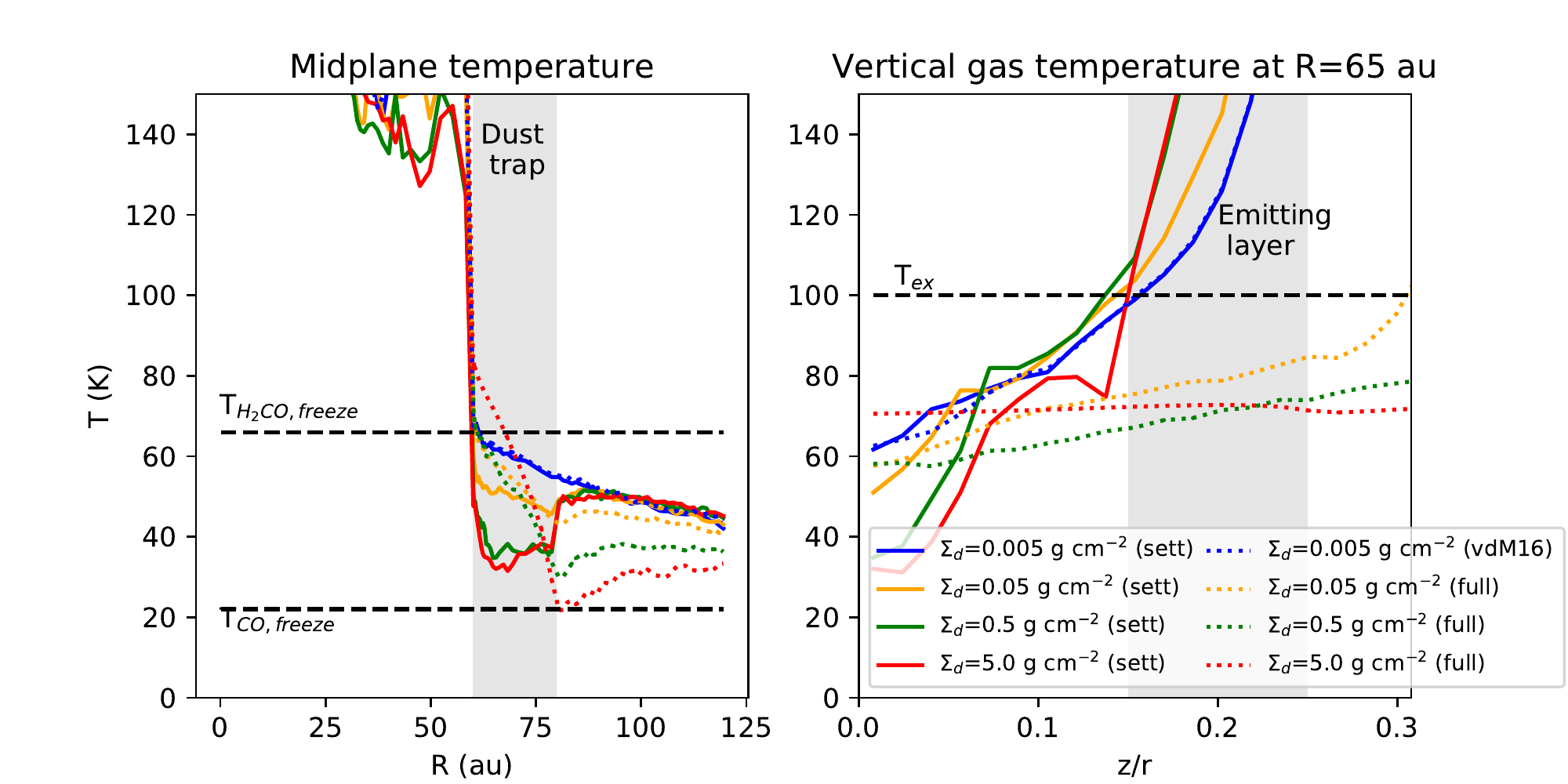}
\includegraphics[width=0.45\textwidth]{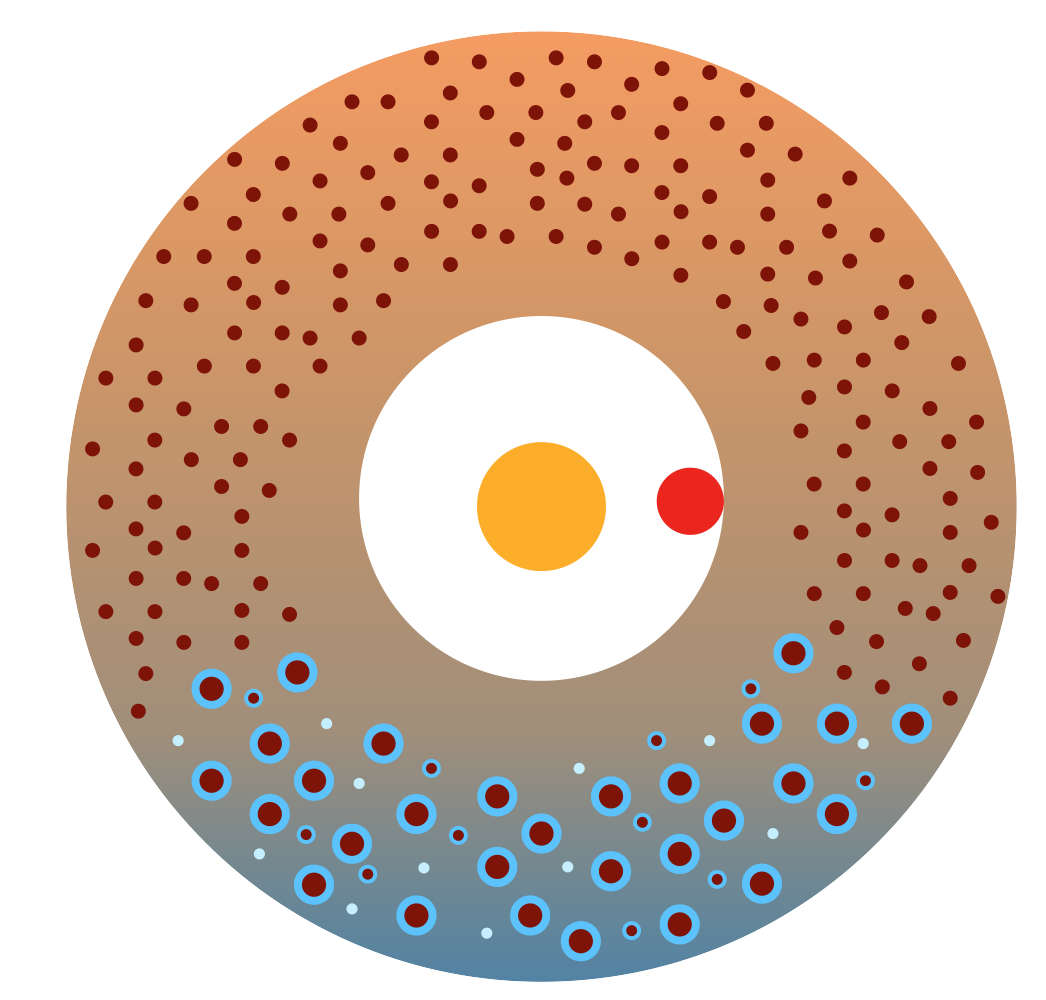}
\includegraphics[width=0.5\textwidth]{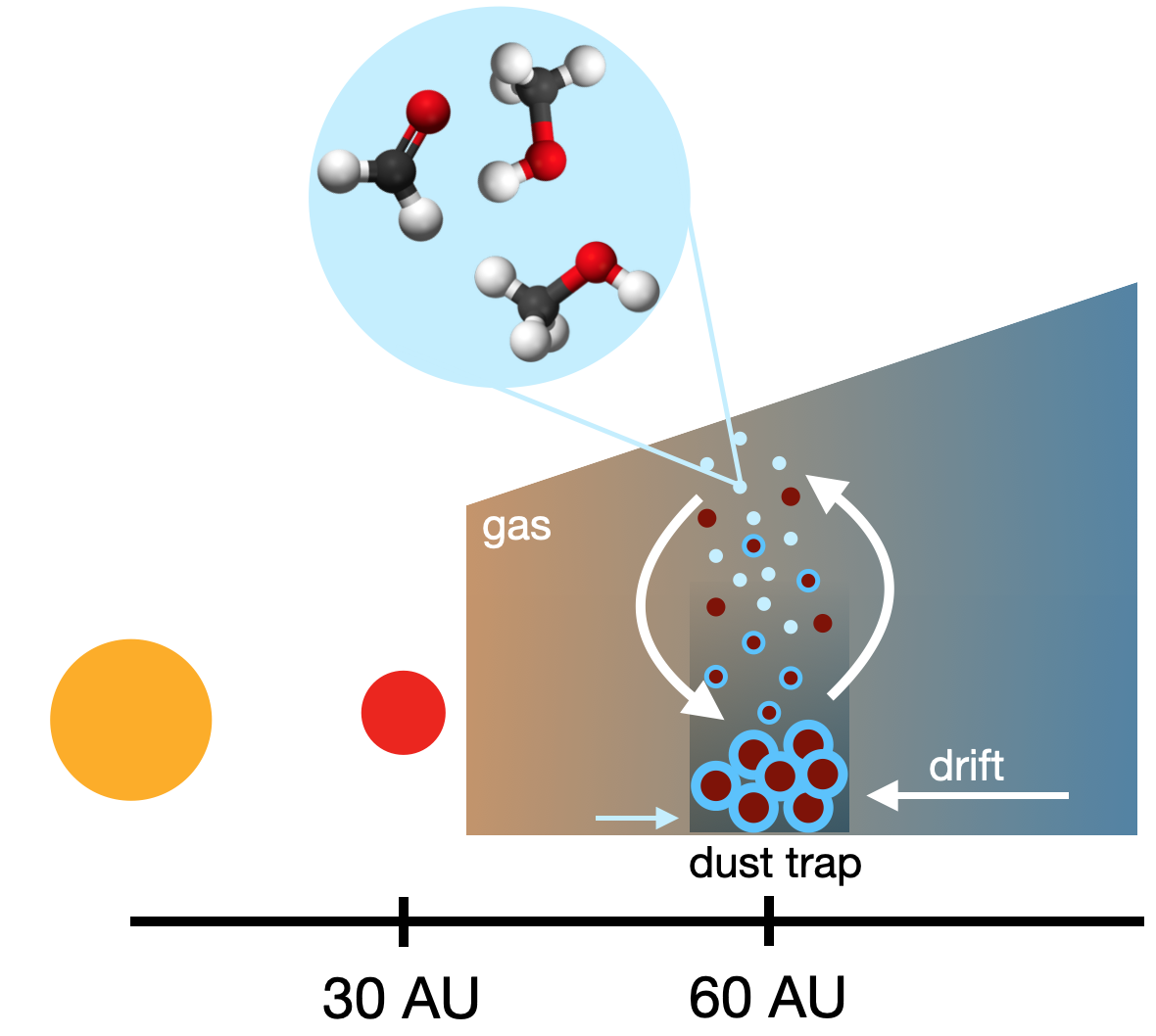}
\caption{{\bf Top left:} Radial dust temperature profiles of the midplane. {\bf Top right:} Vertical gas temperature profiles at 65~au, just inside the dust trap. Both temperature profiles are based on our physical-chemical DALI models with different dust surface densities (Figure \ref{fig:dali},\ref{fig:dali2}). In the midplane, the gas temperature is equal to the dust temperature. The plots demonstrate that the dust trap provides sufficiently low temperatures for a COM ice reservoir in the settled midplane, whereas the temperatures in the emitting layer are sufficiently high to explain the derived excitation temperatures. {\bf Bottom:} Sketch of the proposed scenario for the complex organic chemistry in a dust trap based on this work. The blue-orange gradient indicates the predicted temperature structure in the disk the large arrows the vertical and radial transport of the icy pebbles, and the small arrow the thermal desorption at the inner edge of the dust trap}.
\label{fig:temperature}
\end{figure*}
 
Figure \ref{fig:temperature} shows the midplane temperature profiles and the vertical gas temperature profiles at 65~au (away from the edge), based on the model output (Figure \ref{fig:dali}-\ref{fig:dali2}). The gas temperature is coupled to the dust temperature in the midplane, up to $z/r\lesssim0.15$. Figure \ref{fig:temperature} demonstrates that the midplane temperature may get as low as 40~K in the dust trap, well below the H$_2$CO sublimation temperature of 66~K \citep{Penteado2017}. The temperature is unlikely to be below the CO freezeout temperature of 22~K, so continuous formation of H$_2$CO and CH$_3$OH through CO ice hydrogenation is not possible. However, the dust trap contains a very large reservoir of icy grains. In order to sublimate, icy grains containing H$_2$CO and CH$_3$OH need to be vertically transported to the higher, warmer layers in the disk. The right panel in Figure \ref{fig:temperature} shows that the temperature reaches $>$100~K at a height of $z/r\sim$0.2 for the settled models, in the emitting layer of H$_2$CO and CH$_3$OH. The full models reach temperature of only 80~K, not high enough to explain the observed excitation temperatures, suggesting that the high dust concentration of large dust grains must be settled to the midplane. We note that the gas temperature in the warm molecular layer in IRS48 is estimated as 250-350 K from mid-J rotational CO lines \citep{Fedele2016}, and also that the brightness temperature of the $^{13}$CO 6--5 line  is as high as $\sim$200 K \citep{vanderMarel2016-isot}.

This distribution of icy dust grains and molecules requires efficient vertical transport in the disk which can be achieved by turbulent mixing \citep{Semenov2011}. In addition, the dynamics of the vortex itself may play a role, as both the vertical shear instability \citep{Flock2020} and meridional flows \citep{Meheut2010} in vortices increase the vertical mixing. The full scenario of the ice reservoir, mixing and sublimation is summarized in the bottom panels of Figure \ref{fig:temperature}. 

The morphology of the COM emission suggests that the combination of a concentration of icy dust pebbles and the irradiation of the cavity wall (resulting in a thin layer of thermal sublimation as proposed by \citealt{Cleeves2011}) increase the chemical complexity in IRS~48. It is also important to point out that the COM emission is unlikely to trace an actual azimuthal gas overdensity in the disk: considering the high SNR, the azimuthal contrast would have to be more than a factor of 10, and such a contrast was not detected in the {C$^{18}$O 6-5 isotopologue emission \citep{vanderMarel2016-isot}, which is optically thin considering the ratio of 4.3$\pm$1.4 with the C$^{17}$O flux measured by \citet{Bruderer2014}.

The difference in azimuthal extent between H$_2$CO and CH$_3$OH could be explained by the lower desorption temperature of H$_2$CO compared to CH$_3$OH. This discrepancy cannot be explained by additional gas phase chemistry production, as this would result in H$_2$CO emission along the entire ring. Formation by gas phase chemistry is not excluded but due to the extreme dust trapping the ice sublimation dominates the observable chemistry.

It is unclear whether the high dust-to-gas ratio environment and high COM abundances are unique for IRS~48. \citet{Pegues2020} and \citet{Facchini2021} derive H$_2$CO column densities of $\sim$10$^{12}$~cm$^{-2}$ and low excitation temperatures of 20-30~K for a number of T Tauri disks, whereas for HD100546, the rotational temperature in the inner 50 au was derived as 50-100 K \citep{Booth2021}. All these T Tauri disks have radial dust traps, which also lead to dust concentrations smaller than the beam size of the COM observations. Interestingly, \citet{Pegues2020} find a higher column density of $\sim10^{13}$~cm$^{-2}$ for J1604-2130, the only disk in their sample for which the H$_2$CO emission is more resolved radially, revealing the ring structure also seen in the continuum. As the inferred column densities rely on the assumed emitting area, the column densities in these works are lower limits and may be as high as in IRS~48. However, the high excitation temperatures of the COMs in IRS~48 require more efficient vertical transport, which may be due to the unique vortex properties. 

IRS~48 is the first protoplanetary disk with a clear link between the morphology of the COM emission and the continuum. These results show the importance of taking into account dust traps in chemical disk models in the production of complex organic chemistry, and in spatially resolving COMs in disks for comparison with the dust substructure.

\begin{acknowledgements}
We thank the referee for their thoughtful report, which has improved the clarity of the manuscript.
We would also like to thank Wlad Lyra and Sebastiaan Krijt for useful discussions and Akimasa Kataoka for his help with the reduction of the data. 
N.M. acknowledges support from the Banting Postdoctoral Fellowships program, administered by the Government of Canada.
ALMA is a partnership of ESO (representing its member states), NSF (USA) and NINS (Japan), together with NRC (Canada) and NSC and ASIAA (Taiwan) and KASI (Republic of Korea), in cooperation with the Republic of Chile. The Joint ALMA Observatory is operated by ESO, AUI/ NRAO and NAOJ. This paper makes use of the following ALMA data: 2017.1.00834.S.
\end{acknowledgements}

\vspace{5mm}

\appendix

\bibliographystyle{aa}

\begin{thebibliography}{62}
\expandafter\ifx\csname natexlab\endcsname\relax\def\natexlab#1{#1}\fi

\bibitem[{{Ag{\'u}ndez} {et~al.}(2018){Ag{\'u}ndez}, {Roueff}, {Le Petit}, \&
  {Le Bourlot}}]{Agundez2018}
{Ag{\'u}ndez}, M., {Roueff}, E., {Le Petit}, F., \& {Le Bourlot}, J. 2018,
  \aap, 616, A19

\bibitem[{{Alarc{\'o}n} {et~al.}(2020){Alarc{\'o}n}, {Teague}, {Zhang},
  {Bergin}, \& {Barraza-Alfaro}}]{Alarcon2020}
{Alarc{\'o}n}, F., {Teague}, R., {Zhang}, K., {Bergin}, E.~A., \&
  {Barraza-Alfaro}, M. 2020, \apj, 905, 68

\bibitem[{{Andrews} {et~al.}(2018){Andrews}, {Huang}, {P{\'e}rez}, {Isella},
  {Dullemond}, {Kurtovic}, {Guzm{\'a}n}, {Carpenter}, {Wilner}, {Zhang}, {Zhu},
  {Birnstiel}, {Bai}, {Benisty}, {Hughes}, {{\"O}berg}, \&
  {Ricci}}]{Andrews2018}
{Andrews}, S.~M., {Huang}, J., {P{\'e}rez}, L.~M., {et~al.} 2018, \apjl, 869,
  L41

\bibitem[{{Andrews} {et~al.}(2011){Andrews}, {Wilner}, {Espaillat}, {Hughes},
  {Dullemond}, {McClure}, {Qi}, \& {Brown}}]{Andrews2011}
{Andrews}, S.~M., {Wilner}, D.~J., {Espaillat}, C., {et~al.} 2011, \apj, 732,
  42

\bibitem[{{Barge} \& {Sommeria}(1995)}]{BargeSommeria1995}
{Barge}, P. \& {Sommeria}, J. 1995, \aap, 295, L1

\bibitem[{{Bergin} {et~al.}(2007){Bergin}, {Aikawa}, {Blake}, \& {van
  Dishoeck}}]{Bergin2007}
{Bergin}, E.~A., {Aikawa}, Y., {Blake}, G.~A., \& {van Dishoeck}, E.~F. 2007,
  Protostars and Planets V, 751

\bibitem[{{Boogert} {et~al.}(2015){Boogert}, {Gerakines}, \&
  {Whittet}}]{Boogert2015}
{Boogert}, A.~C.~A., {Gerakines}, P.~A., \& {Whittet}, D. C.~B. 2015, \araa,
  53, 541

\bibitem[{{Booth} {et~al.}(2021){Booth}, {Walsh}, {Terwisscha van Scheltinga},
  {van Dishoeck}, {Ilee}, {Hogerheijde}, {Kama}, \& {Nomura}}]{Booth2021}
{Booth}, A.~S., {Walsh}, C., {Terwisscha van Scheltinga}, J., {et~al.} 2021,
  Nature Astronomy [\eprint[arXiv]{2104.08348}]

\bibitem[{{Brauer} {et~al.}(2008){Brauer}, {Dullemond}, \&
  {Henning}}]{Brauer2008}
{Brauer}, F., {Dullemond}, C.~P., \& {Henning}, T. 2008, \aap, 480, 859

\bibitem[{{Bruderer}(2013)}]{Bruderer2013}
{Bruderer}, S. 2013, \aap, 559, A46

\bibitem[{{Bruderer} {et~al.}(2014){Bruderer}, {van der Marel}, {van Dishoeck},
  \& {van Kempen}}]{Bruderer2014}
{Bruderer}, S., {van der Marel}, N., {van Dishoeck}, E.~F., \& {van Kempen},
  T.~A. 2014, \aap, 562, A26

\bibitem[{{Bruderer} {et~al.}(2012){Bruderer}, {van Dishoeck}, {Doty}, \&
  {Herczeg}}]{Bruderer2012}
{Bruderer}, S., {van Dishoeck}, E.~F., {Doty}, S.~D., \& {Herczeg}, G.~J. 2012,
  \aap, 541, A91

\bibitem[{{Cleeves} {et~al.}(2011){Cleeves}, {Bergin}, {Bethell}, {Calvet},
  {Fogel}, {Sauter}, \& {Wolf}}]{Cleeves2011}
{Cleeves}, L.~I., {Bergin}, E.~A., {Bethell}, T.~J., {et~al.} 2011, \apjl, 743,
  L2

\bibitem[{{Cridland} {et~al.}(2017){Cridland}, {Pudritz}, \&
  {Birnstiel}}]{Cridland2017}
{Cridland}, A.~J., {Pudritz}, R.~E., \& {Birnstiel}, T. 2017, \mnras, 465, 3865

\bibitem[{{Dutrey} {et~al.}(1997){Dutrey}, {Guilloteau}, \&
  {Guelin}}]{Dutrey1997}
{Dutrey}, A., {Guilloteau}, S., \& {Guelin}, M. 1997, \aap, 317, L55

\bibitem[{{Ehrenfreund} \& {Charnley}(2000)}]{Ehrenfreund2000}
{Ehrenfreund}, P. \& {Charnley}, S.~B. 2000, \araa, 38, 427

\bibitem[{{Facchini} {et~al.}(2021){Facchini}, {Teague}, {Bae}, {Benisty},
  {Keppler}, \& {Isella}}]{Facchini2021}
{Facchini}, S., {Teague}, R., {Bae}, J., {et~al.} 2021, arXiv e-prints,
  arXiv:2101.08369

\bibitem[{{Fedele} {et~al.}(2016){Fedele}, {van Dishoeck}, {Kama}, {Bruderer},
  \& {Hogerheijde}}]{Fedele2016}
{Fedele}, D., {van Dishoeck}, E.~F., {Kama}, M., {Bruderer}, S., \&
  {Hogerheijde}, M.~R. 2016, \aap, 591, A95

\bibitem[{{Flock} {et~al.}(2020){Flock}, {Turner}, {Nelson}, {Lyra}, {Manger},
  \& {Klahr}}]{Flock2020}
{Flock}, M., {Turner}, N.~J., {Nelson}, R.~P., {et~al.} 2020, \apj, 897, 155

\bibitem[{{Foreman-Mackey} {et~al.}(2013){Foreman-Mackey}, {Hogg}, {Lang}, \&
  {Goodman}}]{DFM2013}
{Foreman-Mackey}, D., {Hogg}, D.~W., {Lang}, D., \& {Goodman}, J. 2013, \pasp,
  125, 306

\bibitem[{{Francis} \& {van der Marel}(2020)}]{Francis2020}
{Francis}, L. \& {van der Marel}, N. 2020, \apj, 892, 111

\bibitem[{{Fuchs} {et~al.}(2009){Fuchs}, {Cuppen}, {Ioppolo}, {Romanzin},
  {Bisschop}, {Andersson}, {van Dishoeck}, \& {Linnartz}}]{Fuchs2009}
{Fuchs}, G.~W., {Cuppen}, H.~M., {Ioppolo}, S., {et~al.} 2009, \aap, 505, 629

\bibitem[{{Gaia Collaboration} {et~al.}(2018){Gaia Collaboration}, {Brown},
  {Vallenari}, {Prusti}, {de Bruijne}, {Babusiaux}, {Bailer-Jones}, {Biermann},
  {Evans}, {Eyer}, \& et~al.}]{Gaia2018}
{Gaia Collaboration}, {Brown}, A.~G.~A., {Vallenari}, A., {et~al.} 2018, \aap,
  616, A1

\bibitem[{{Garrod} {et~al.}(2006){Garrod}, {Park}, {Caselli}, \&
  {Herbst}}]{Garrod2006}
{Garrod}, R., {Park}, I.~H., {Caselli}, P., \& {Herbst}, E. 2006, Faraday
  Discussions, 133, 51

\bibitem[{{Geers} {et~al.}(2007){Geers}, {Pontoppidan}, {van Dishoeck},
  {Dullemond}, {Augereau}, {Mer{\'{\i}}n}, {Oliveira}, \& {Pel}}]{Geers2007}
{Geers}, V.~C., {Pontoppidan}, K.~M., {van Dishoeck}, E.~F., {et~al.} 2007,
  \aap, 469, L35

\bibitem[{{Hama} {et~al.}(2018){Hama}, {Kouchi}, \& {Watanabe}}]{Hama2018}
{Hama}, T., {Kouchi}, A., \& {Watanabe}, N. 2018, \apjl, 857, L13

\bibitem[{{Herbst} \& {van Dishoeck}(2009)}]{Herbst2009}
{Herbst}, E. \& {van Dishoeck}, E.~F. 2009, \araa, 47, 427

\bibitem[{{Kahane} {et~al.}(1984){Kahane}, {Frerking}, {Langer}, {Encrenas}, \&
  {Lucas}}]{Kahane1984}
{Kahane}, C., {Frerking}, M.~A., {Langer}, W.~D., {Encrenas}, P., \& {Lucas},
  R. 1984, \aap, 137, 211

\bibitem[{{Krijt} {et~al.}(2020){Krijt}, {Bosman}, {Zhang}, {Schwarz},
  {Ciesla}, \& {Bergin}}]{Krijt2020}
{Krijt}, S., {Bosman}, A.~D., {Zhang}, K., {et~al.} 2020, \apj, 899, 134

\bibitem[{{Lee} {et~al.}(2019){Lee}, {Lee}, {Baek}, {Aikawa}, {Cieza}, {Yoon},
  {Herczeg}, {Johnstone}, \& {Casassus}}]{Lee2019}
{Lee}, J.-E., {Lee}, S., {Baek}, G., {et~al.} 2019, Nature Astronomy, 3, 314

\bibitem[{{Loomis} {et~al.}(2018){Loomis}, {{\"O}berg}, {Andrews}, {Walsh},
  {Czekala}, {Huang}, \& {Rosenfeld}}]{Loomis2018}
{Loomis}, R.~A., {{\"O}berg}, K.~I., {Andrews}, S.~M., {et~al.} 2018, \aj, 155,
  182

\bibitem[{{Mangum} \& {Wootten}(1993)}]{Mangum1993}
{Mangum}, J.~G. \& {Wootten}, A. 1993, \apjs, 89, 123

\bibitem[{{Meheut} {et~al.}(2010){Meheut}, {Casse}, {Varniere}, \&
  {Tagger}}]{Meheut2010}
{Meheut}, H., {Casse}, F., {Varniere}, P., \& {Tagger}, M. 2010, \aap, 516, A31

\bibitem[{{Mulders} {et~al.}(2011){Mulders}, {Waters}, {Dominik}, {Sturm},
  {Bouwman}, {Min}, {Verhoeff}, {Acke}, {Augereau}, {Evans}, {Henning},
  {Meeus}, \& {Olofsson}}]{Mulders2011}
{Mulders}, G.~D., {Waters}, L.~B.~F.~M., {Dominik}, C., {et~al.} 2011, \aap,
  531, A93

\bibitem[{{{\"O}berg} \& {Bergin}(2021)}]{Oberg2020}
{{\"O}berg}, K.~I. \& {Bergin}, E.~A. 2021, \physrep, 893, 1

\bibitem[{{{\"O}berg} {et~al.}(2015){{\"O}berg}, {Guzm{\'a}n}, {Furuya}, {Qi},
  {Aikawa}, {Andrews}, {Loomis}, \& {Wilner}}]{Oberg2015}
{{\"O}berg}, K.~I., {Guzm{\'a}n}, V.~V., {Furuya}, K., {et~al.} 2015, \nat,
  520, 198

\bibitem[{{{\"O}berg} {et~al.}(2010){{\"O}berg}, {Qi}, {Fogel}, {Bergin},
  {Andrews}, {Espaillat}, {van Kempen}, {Wilner}, \& {Pascucci}}]{Oberg2010}
{{\"O}berg}, K.~I., {Qi}, C., {Fogel}, J.~K.~J., {et~al.} 2010, \apj, 720, 480

\bibitem[{{Ohashi} {et~al.}(2020){Ohashi}, {Kataoka}, {van der Marel}, {Hull},
  {Dent}, {Pohl}, {Pinilla}, {van Dishoeck}, \& {Henning}}]{Ohashi2020}
{Ohashi}, S., {Kataoka}, A., {van der Marel}, N., {et~al.} 2020, \apj, 900, 81

\bibitem[{{Pegues} {et~al.}(2020){Pegues}, {{\"O}berg}, {Bergner}, {Loomis},
  {Qi}, {Le Gal}, {Cleeves}, {Guzm{\'a}n}, {Huang}, {J{\o}rgensen}, {Andrews},
  {Blake}, {Carpenter}, {Schwarz}, {Williams}, \& {Wilner}}]{Pegues2020}
{Pegues}, J., {{\"O}berg}, K.~I., {Bergner}, J.~B., {et~al.} 2020, \apj, 890,
  142

\bibitem[{{Penteado} {et~al.}(2017){Penteado}, {Walsh}, \&
  {Cuppen}}]{Penteado2017}
{Penteado}, E.~M., {Walsh}, C., \& {Cuppen}, H.~M. 2017, \apj, 844, 71

\bibitem[{{P{\'e}rez} {et~al.}(2014){P{\'e}rez}, {Isella}, {Carpenter}, \&
  {Chandler}}]{Perez2014}
{P{\'e}rez}, L.~M., {Isella}, A., {Carpenter}, J.~M., \& {Chandler}, C.~J.
  2014, \apjl, 783, L13

\bibitem[{{Pinilla} {et~al.}(2012{\natexlab{a}}){Pinilla}, {Benisty}, \&
  {Birnstiel}}]{Pinilla2012b}
{Pinilla}, P., {Benisty}, M., \& {Birnstiel}, T. 2012{\natexlab{a}}, \aap, 545,
  A81

\bibitem[{{Pinilla} {et~al.}(2012{\natexlab{b}}){Pinilla}, {Birnstiel},
  {Ricci}, {Dullemond}, {Uribe}, {Testi}, \& {Natta}}]{Pinilla2012a}
{Pinilla}, P., {Birnstiel}, T., {Ricci}, L., {et~al.} 2012{\natexlab{b}}, \aap,
  538, A114

\bibitem[{{Podio} {et~al.}(2020){Podio}, {Garufi}, {Codella}, {Fedele},
  {Bianchi}, {Bacciotti}, {Ceccarelli}, {Favre}, {Mercimek}, {Rygl}, \&
  {Testi}}]{Podio2020}
{Podio}, L., {Garufi}, A., {Codella}, C., {et~al.} 2020, \aap, 642, L7

\bibitem[{{Rabli} \& {Flower}(2010)}]{Rabli2010}
{Rabli}, D. \& {Flower}, D.~R. 2010, \mnras, 406, 95

\bibitem[{{Sch{\"o}ier} {et~al.}(2005){Sch{\"o}ier}, {van der Tak}, {van
  Dishoeck}, \& {Black}}]{Schoier2005}
{Sch{\"o}ier}, F.~L., {van der Tak}, F.~F.~S., {van Dishoeck}, E.~F., \&
  {Black}, J.~H. 2005, \aap, 432, 369

\bibitem[{{Semenov} \& {Wiebe}(2011)}]{Semenov2011}
{Semenov}, D. \& {Wiebe}, D. 2011, \apjs, 196, 25

\bibitem[{{Terwisscha van Scheltinga} {et~al.}(2021){Terwisscha van
  Scheltinga}, {Hogerheijde}, {Cleeves}, {Loomis}, {Walsh}, {{\"O}berg},
  {Bergin}, {Bergner}, {Blake}, {Calahan}, {Cazzoletti}, {van Dishoeck},
  {Guzm{\'a}n}, {Huang}, {Kama}, {Qi}, {Teague}, \& {Wilner}}]{TwS2021}
{Terwisscha van Scheltinga}, J., {Hogerheijde}, M.~R., {Cleeves}, L.~I.,
  {et~al.} 2021, \apj, 906, 111

\bibitem[{{Thi} {et~al.}(2004){Thi}, {van Zadelhoff}, \& {van
  Dishoeck}}]{Thi2004}
{Thi}, W.-F., {van Zadelhoff}, G.-J., \& {van Dishoeck}, E.~F. 2004, \aap, 425,
  955

\bibitem[{{van der Marel} {et~al.}(2015){van der Marel}, {Pinilla}, {Tobin},
  {van Kempen}, {Andrews}, {Ricci}, \& {Birnstiel}}]{vanderMarel2015vla}
{van der Marel}, N., {Pinilla}, P., {Tobin}, J., {et~al.} 2015, \apjl, 810, L7

\bibitem[{{van der Marel} {et~al.}(2016){van der Marel}, {van Dishoeck},
  {Bruderer}, {Andrews}, {Pontoppidan}, {Herczeg}, {van Kempen}, \&
  {Miotello}}]{vanderMarel2016-isot}
{van der Marel}, N., {van Dishoeck}, E.~F., {Bruderer}, S., {et~al.} 2016,
  \aap, 585, A58

\bibitem[{{van der Marel} {et~al.}(2013){van der Marel}, {van Dishoeck},
  {Bruderer}, {Birnstiel}, {Pinilla}, {Dullemond}, {van Kempen}, {Schmalzl},
  {Brown}, {Herczeg}, {Mathews}, \& {Geers}}]{vanderMarel2013}
{van der Marel}, N., {van Dishoeck}, E.~F., {Bruderer}, S., {et~al.} 2013,
  Science, 340, 1199

\bibitem[{{van der Marel} {et~al.}(2014){van der Marel}, {van Dishoeck},
  {Bruderer}, \& {van Kempen}}]{vanderMarel2014}
{van der Marel}, N., {van Dishoeck}, E.~F., {Bruderer}, S., \& {van Kempen},
  T.~A. 2014, \aap, 563, A113

\bibitem[{{van der Tak} {et~al.}(2007){van der Tak}, {Black}, {Sch{\"o}ier},
  {Jansen}, \& {van Dishoeck}}]{vanderTak2007}
{van der Tak}, F.~F.~S., {Black}, J.~H., {Sch{\"o}ier}, F.~L., {Jansen}, D.~J.,
  \& {van Dishoeck}, E.~F. 2007, \aap, 468, 627

\bibitem[{{van Dishoeck} {et~al.}(1993){van Dishoeck}, {Blake}, {Draine}, \&
  {Lunine}}]{vanDishoeck1993}
{van Dishoeck}, E.~F., {Blake}, G.~A., {Draine}, B.~T., \& {Lunine}, J.~I.
  1993, in Protostars and Planets III, ed. E.~H. {Levy} \& J.~I. {Lunine}, 163

\bibitem[{{van Dishoeck} {et~al.}(1995){van Dishoeck}, {Blake}, {Jansen}, \&
  {Groesbeck}}]{vanDishoeck1995}
{van Dishoeck}, E.~F., {Blake}, G.~A., {Jansen}, D.~J., \& {Groesbeck}, T.~D.
  1995, \apj, 447, 760

\bibitem[{{van 't Hoff} {et~al.}(2018){van 't Hoff}, {Tobin}, {Trapman},
  {Harsono}, {Sheehan}, {Fischer}, {Megeath}, \& {van Dishoeck}}]{vantHoff2018}
{van 't Hoff}, M. L.~R., {Tobin}, J.~J., {Trapman}, L., {et~al.} 2018, \apjl,
  864, L23

\bibitem[{{Walsh} {et~al.}(2016){Walsh}, {Juh{\'a}sz}, {Meeus}, {Dent}, {Maud},
  {Aikawa}, {Millar}, \& {Nomura}}]{Walsh2016}
{Walsh}, C., {Juh{\'a}sz}, A., {Meeus}, G., {et~al.} 2016, \apj, 831, 200

\bibitem[{{Walsh} {et~al.}(2014){Walsh}, {Millar}, {Nomura}, {Herbst}, {Widicus
  Weaver}, {Aikawa}, {Laas}, \& {Vasyunin}}]{Walsh2014}
{Walsh}, C., {Millar}, T.~J., {Nomura}, H., {et~al.} 2014, \aap, 563, A33

\bibitem[{{Watanabe} \& {Kouchi}(2002)}]{Watanabe2002}
{Watanabe}, N. \& {Kouchi}, A. 2002, \apjl, 571, L173

\bibitem[{{Weidenschilling}(1977)}]{Weidenschilling1977}
{Weidenschilling}, S.~J. 1977, \mnras, 180, 57

\bibitem[{{Wiesenfeld} \& {Faure}(2013)}]{Wiesenfeld2013}
{Wiesenfeld}, L. \& {Faure}, A. 2013, \mnras, 432, 2573

\end{thebibliography}

\appendix
\onecolumn

\section{Spectra and intensity maps}
\label{sct:spectra}
This section contains additional figures of the data.

\begin{figure*}[!ht]
    \centering
    \includegraphics{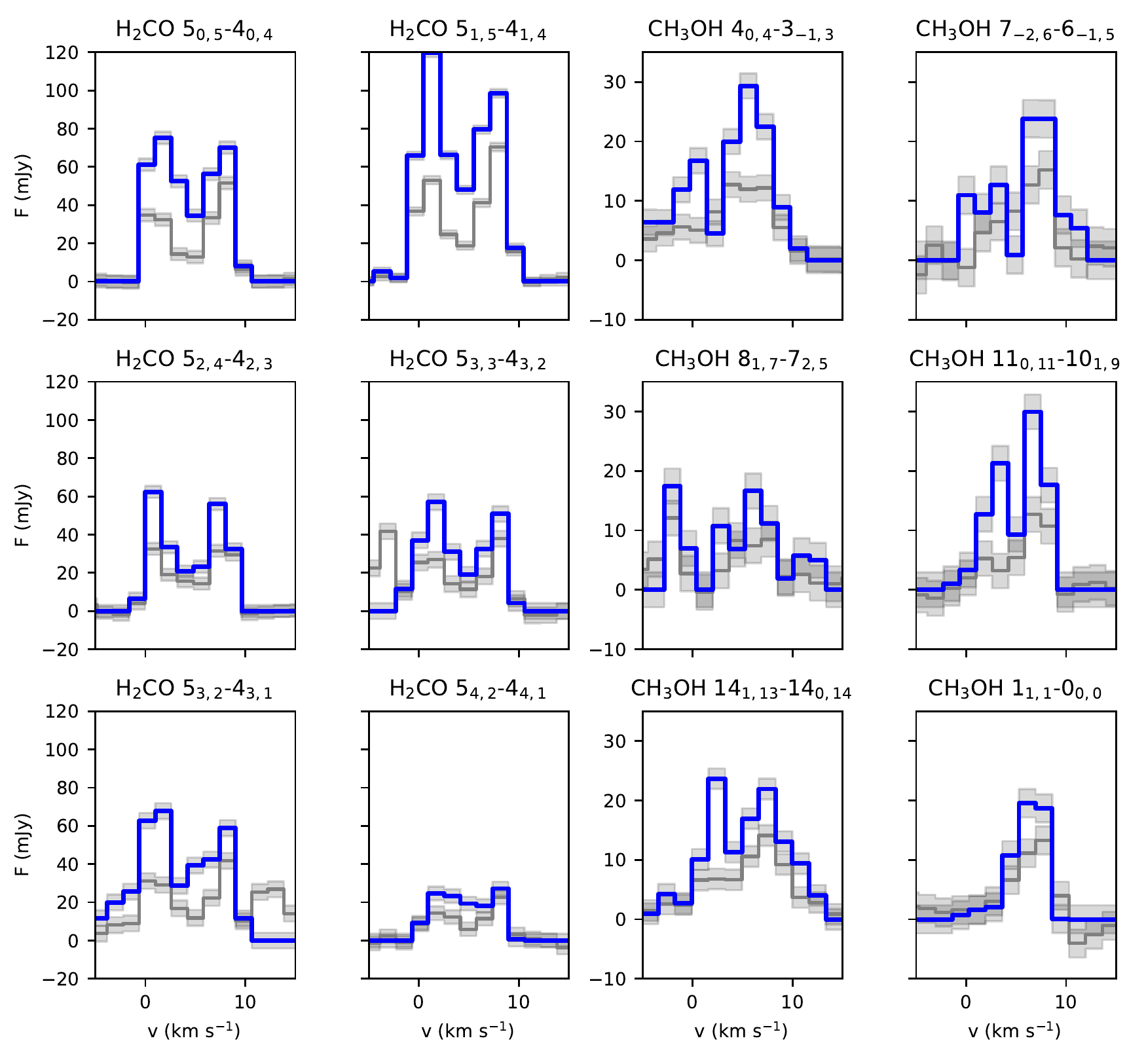}
    \caption{H$_2$CO and CH$_3$OH spectra, integrated over the central area of the dust trap using Keplerian masking (blue) and extracted from a rectangular box (grey).  The grey shades indicate the noise levels in the spectra. The spectra are ordered by increasing $E_u$, following the values reported in Table \ref{tab:lines}.}
    \label{fig:spectra}
\end{figure*}

\newpage

\begin{figure*}[!ht]
    \centering
    \includegraphics{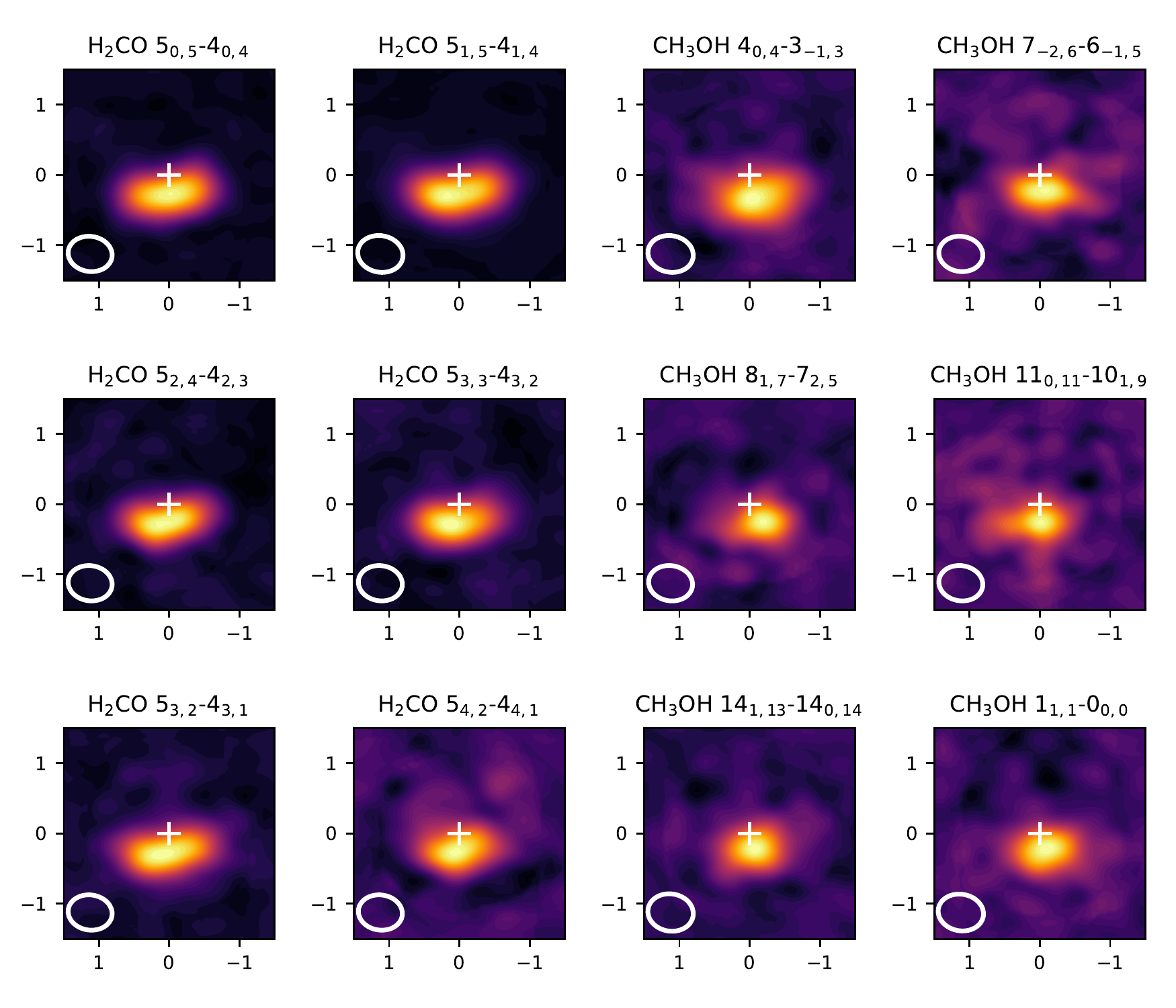}
    \caption{H$_2$CO and CH$_3$OH naturally weighted zero moment maps using Keplerian masking. The plus indicates the position of the star and the beam is shown in the lower left of each map.}
    \label{fig:mommaps}
\end{figure*}

\begin{figure*}[!ht]
    \centering
    \includegraphics[width=0.6\textwidth]{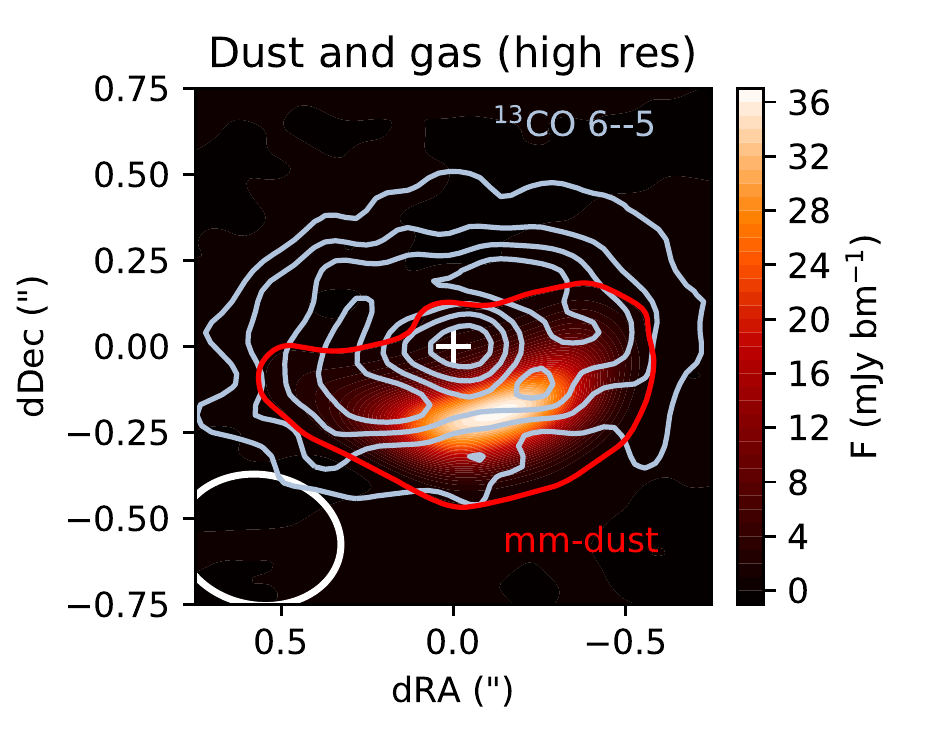}
    \caption{High resolution image of the $^{13}$CO 6--5 zero moment map (blue) and the 366 GHz continuum (red) at the original 0.18$\times$0.14" resolution. The contours of the $^{13}$CO show the 20,40,60,80\% of the peak while the contours of the continuum indicate the 5$\sigma$ level. This image demonstrates that the gas traces a full disk ring whereas the mm-dust grains are concentrated in the southern part of the disk.}
    \label{fig:highres}
\end{figure*}

\newpage
\section{Rotational diagram analysis}
\label{sec:mcmc}
This section contains more details on the analysis of the rotational diagram from Figure \ref{fig:rotdiagram}. Figure \ref{fig:posteriors} shows the corner plots showing the posterior distributions and covariances of the fit containing all line transitions, confirming the fit has converged. The covariance is similar to previous studies of rotational diagrams of COMs in protoplanetary disks \citep[e.g.][]{Loomis2018}.

\begin{figure*}[!ht]
\centering
\includegraphics[width=0.42\textwidth,trim=10 30 10 10]{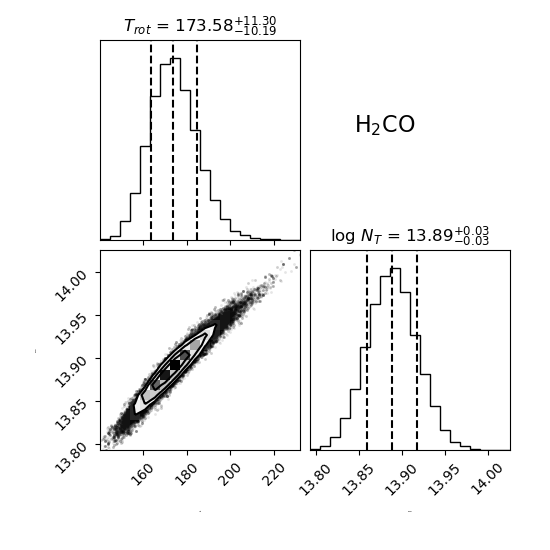}
\includegraphics[width=0.42\textwidth,trim=10 30 10 10]{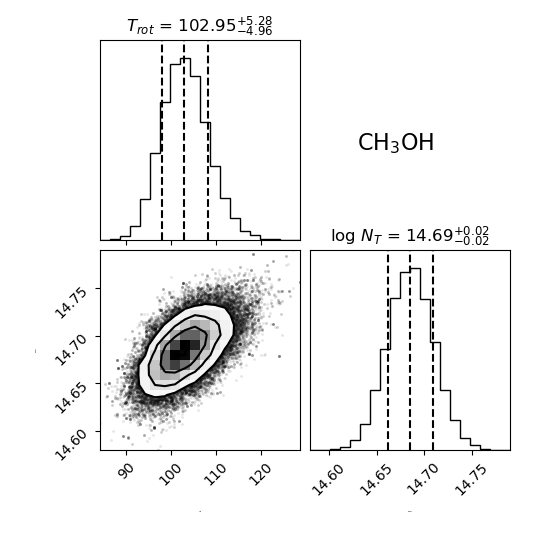}
\caption{Posterior distributions of the column density and rotational temperature based on our optically thin line intensities of H$_2$CO (left) and CH$_3$OH (right). The best fit is shown in Figure \ref{fig:rotdiagram}.}
\label{fig:posteriors}
\end{figure*}

Second, we estimate the ortho-to-para ratio of H$_2$CO by fitting the ortho and para lines separately, without the ortho-to-para correction of 3 in the degeneracies as used in Figure \ref{fig:rotdiagram}. The new rotational diagram with best fits and the posteriors are shown in Figure \ref{fig:rotdiagramopr}. The ortho lines have a best fit $T_{\rm rot}$ of 193$^{+18}_{-15}$ K and $N_T=7.2\pm0.7\cdot10^{13}$ cm$^{-2}$ and the para lines a $T_{\rm rot}$ of 257$^{+57}_{-39}$ K and $N_T=5.9\pm1.7\cdot10^{13}$ cm$^{-2}$. This implies an ortho-to-para ratio of 1.2$\pm$0.4, which is well below the default value of 3. The lower value could indicate an ice formation rather than gas formation route \citep[e.g][]{TwS2021}, which is  consistent with our proposed scenario. However, if the optical depth of the ortho lines is underestimated (if the emitting area is smaller than assumed) this could also explain the lower value. Furthermore, for H$_2$O it has been shown that the ortho-to-para ratio is reset after desorption from the ices \citep{Hama2018} and if the same is true for H$_2$CO, the ortho-to-para ratio does not provide information about the formation origin. Regardless of the precise scenario, a low ortho-to-para ratio would imply a very cold formation location of only $\sim$10 K \citep{Kahane1984}, much lower than the current dust trap conditions.

\begin{figure}[!ht]
\centering
\includegraphics[width=0.45\textwidth]{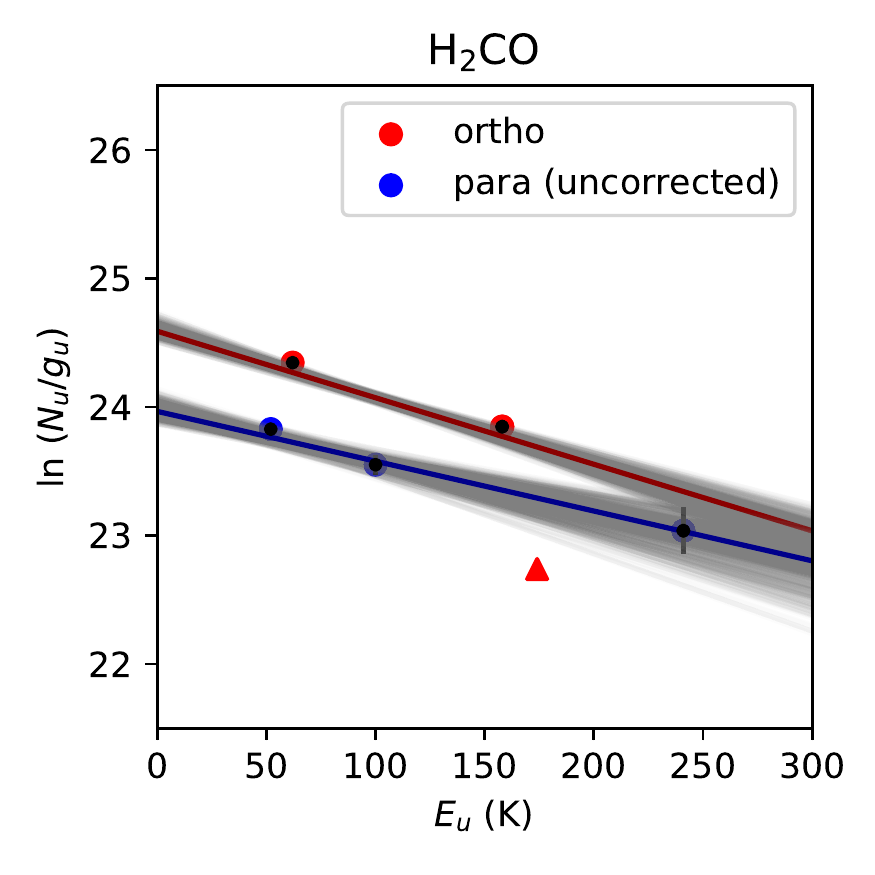}\\
\includegraphics[width=0.45\textwidth]{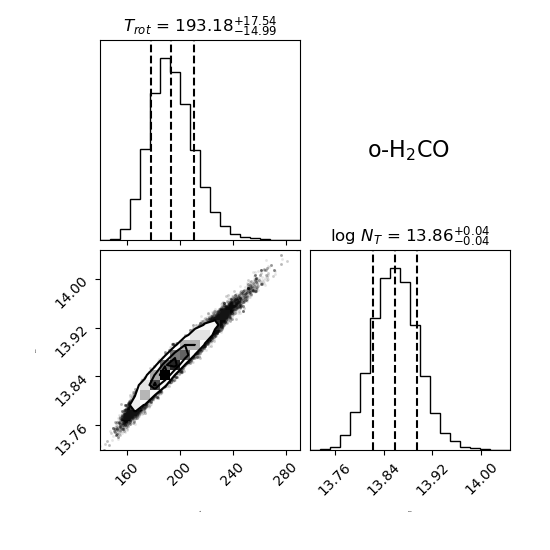}
\includegraphics[width=0.45\textwidth]{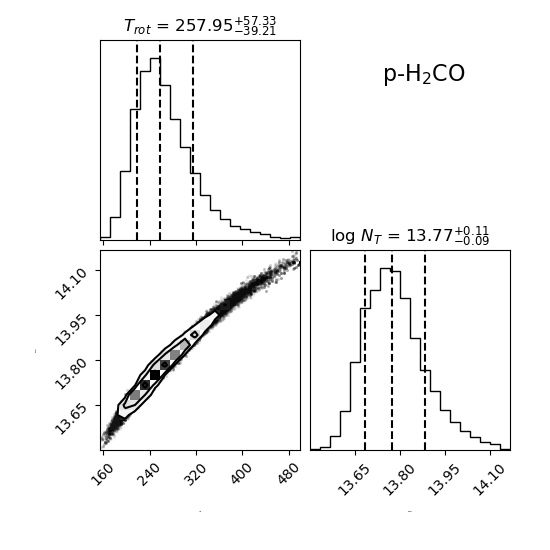}
\caption{Rotational diagram of H$_2$CO, similar to Figure \ref{fig:rotdiagram}, but without the default ortho-to-para correction of 3 for the para lines, with separate fits to the ortho and para lines. The bottom panels show the posterior distributions of the best fits as in Figure \ref{fig:posteriors}.}
\label{fig:rotdiagramopr}
\end{figure}

\section{Temperature structure}
\label{sct:temperature}
This section describes our analysis of the temperature structure and line ratios of IRS~48. We set up a series of physical-chemical models using DALI \citep{Bruderer2012,Bruderer2013} to compute the gas and dust temperature structure from the heating-cooling balance of gas and dust, following the parametrized model of the gas and dust surface density from \citet{vanderMarel2016-isot} that is consistent with the CO 6--5 isotopologue data presented in that work. The dust surface density was not explicitly fit in that model, and with a gas-to-dust ratio of 20 used in that work the dust surface density at 60 au is $\Sigma_d\sim$0.005 g cm$^{-2}$. As \citet{Ohashi2020} derived a much higher dust surface density of $\Sigma_d\sim$2-8 g cm$^{-2}$, we explore the effect of the dust surface density on the radial and vertical temperature structure, using $\Sigma_d\sim$0.05, 0.5 and 5.0 g cm$^{-2}$ between 60 and 80 au. Two sets of models are run: one with an increase of the large grains in the midplane with a larger degree of settling (settled models) and second with an increase of all grains throughout the full dust column (full models). The vertical density structure in DALI is fixed to a Gaussian profile with different scale heights for the large and small grains to represent the settling, following \citet{Andrews2011}:

\begin{align}
\rho_l &= \frac{\Sigma_l}{\sqrt{2\pi}r\chi{h}}\exp\left[-\frac{1}{2}\left(\frac{\pi/2-\theta}{\chi{h}}\right)^2\right] \\
\rho_s &= \frac{\Sigma_s}{\sqrt{2\pi}rh}\exp\left[-\frac{1}{2}\left(\frac{\pi/2-\theta}{h}\right)^2\right] 
\end{align}
with $\rho_l$ and $\rho_s$ the dust density of large and small grains, respectively, $h=h_c(r/r_c)^\psi$ the scale height, $\chi$ the settling parameter and $\theta$ the vertical latitude coordinate measured from the pole. The large grain population contains dust grains from 0.005 $\mu$m to 1 mm and the small grain population contains dust grains from 0.005 - 1 $\mu$m by convention. In the full disk model, the settling degree $\chi$ is set to the default value of 0.2, while in the settled model $\chi$ is set to 0.1. The radial structure inside 60 au contains the inner dust disk (between $r_{\rm sub}$ and $r_{\rm gap}$), the gas cavity $r_{\rm cavgas}$ and dust cavity $r_{\rm cavdust}$, which are the same as in \citet{vanderMarel2016-isot} and we refer for more details to that work. All model parameter values are listed in Table \ref{tbl:model} and the results are shown in Figure \ref{fig:dali} and \ref{fig:dali2}.

\begin{table}[!ht]
\centering
\caption{Model parameters of gas and dust surface density}
\label{tbl:model}
\begin{tabular}{ll|l|l}
\hline
Property&Parameter &Settled models&Full models\\
\hline
Surface density&$\Sigma_{\rm 60 au,gas}$ (g cm$^{-2}$)&0.25&0.25\\
&$\Sigma_{\rm 60 au, dust}$ (g cm$^{-2}$)&0.005,0.05,0.5,5.0&0.005,0.05,0.5,5.0\\
&&(large grains only)&(all grains)\\
Dust settling&$\chi$&0.1&0.2\\
\hline
Vertical structure&$h_c$&0.14&0.14\\
&$\psi$&0.22&0.22\\
\hline

Radial structure&$r_c$ (au)&60&60\\
&$r_{\rm sub}$ (au)&0.4&0.4\\
&$r_{\rm gap}$ (au)&1&1\\
&$r_{\rm cavgas}$ (au)&25&25\\
&$r_{\rm cavdust}$ (au)&60&60\\
&$r_{\rm out}$ (au)&90&90\\
\hline
\end{tabular}
\end{table}

Second, we estimate the expected flux ratios for both optically thick and optically thin emission of the H$_2$CO and CH$_3$OH lines studied in this work. This is parametrized by setting an abundance of 10$^{-7}$ and 10$^{-9}$ (optically thick and thin, respectively) for both molecules in the regions in the disk where the extinction $A_V>$1 (shielding for photodissociation), radius $r>$60 au (in the dust trap) and the dust temperature $T_{\rm dust}>T_{\rm subl}$, with $T_{\rm subl}$=66 K for H$_2$CO and 100 K for CH$_3$OH. These thresholds set the region where the molecule is expected to be in the gas phase. These regions are well above the regime where the majority of the large grains are located due to the settling. In the rest of the disk, these molecular abundances are set to 10$^{-12}$. This leads to specific emitting layers in the disk where H$_2$CO and CH$_3$OH are located, which are used to raytrace the lines to compute line ratios which can be compared with the data. The absolute fluxes are less relevant as the real molecular layers are likely much more complex. Also, as DALI is an axisymmetric model, the azimuthal structure is not constrained in this model. 

\begin{figure*}[!ht]
    \centering
    \includegraphics[width=\textwidth]{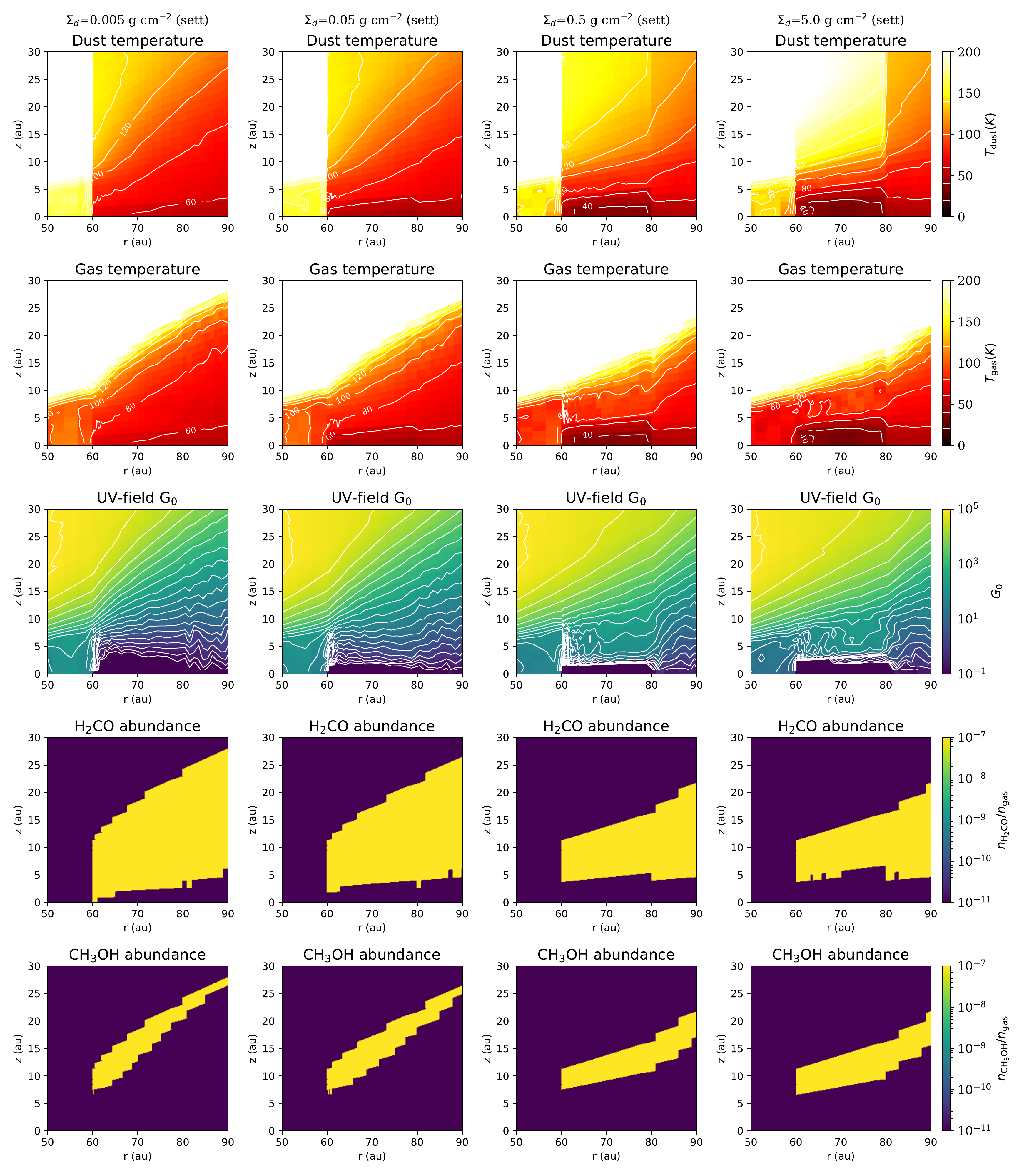}
    \caption{Temperature structure of the IRS~48 disk as computed by DALI, using the gas and dust density profile derived by \citet{vanderMarel2016-isot} for the settled models (dust density increase of the large grains in the midplane)}. The columns show the influence of the assumed dust surface density (or dust-to-gas ratio, as the gas surface density is set constant) on the UV field and  gas temperature. The white contours in the temperature plots indicate steps of 20 K.
    \label{fig:dali}
\end{figure*}

\begin{figure*}[!ht]
    \centering
    \includegraphics[width=\textwidth]{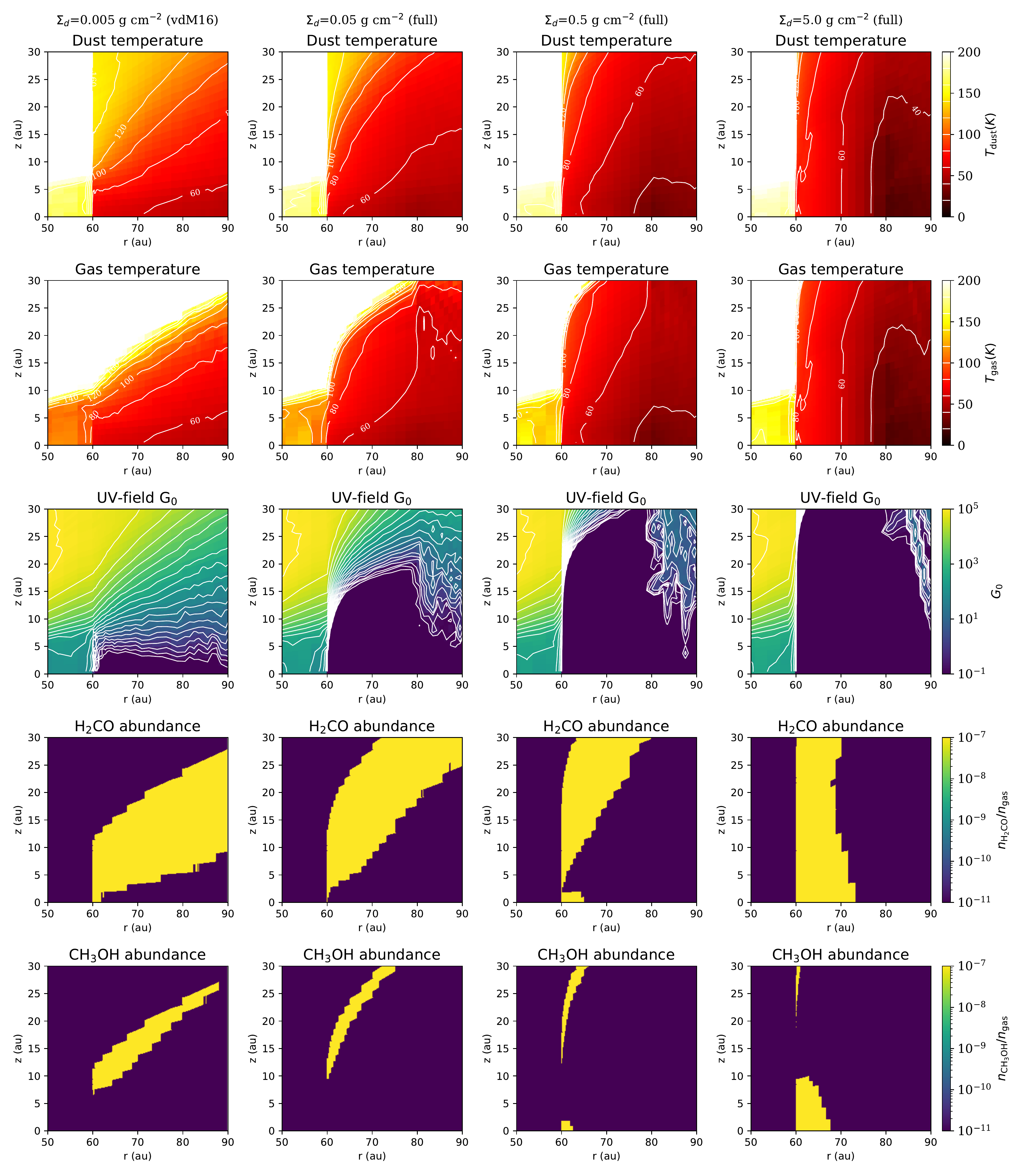}
    \caption{Temperature structure of the IRS~48 disk as computed by DALI, using the gas and dust density profile derived by \citet{vanderMarel2016-isot} for the full models (dust increase across the column). The columns show the influence of the assumed dust surface density (or dust-to-gas ratio, as the gas surface density is set constant) on the UV field and  gas temperature. The white contours in the temperature plots indicate steps of 20 K.}
    \label{fig:dali2}
\end{figure*}

The models show that the dust and gas temperature drop when the dust density is increased due to the stronger suppression of the UV field. This drop remains limited to the midplane when the dust density is only increased in that region compared to an increase throughout the full column. This is illustrated directly in Figure \ref{fig:temperature}. The models show that the emitting layer shifts upward in the disk for higher dust densities when the dust is distributed throughout the disk ('full') due to the sublimation temperature requirement. On the upper end, the emitting layer is constrained by the extinction requirement, which means that the layer becomes thinner in the high dust density models. In the most extreme case of $\Sigma_d$=5 g cm$^{-2}$ an additional emitting layer appears in the midplane as the dust becomes fully optically thick at the dust edge, leading to a strong vertical increase in temperature. 

The raytraced line fluxes reproduce the H$_2$CO fluxes reasonably well for the settled models for the 10$^{-7}$ abundance, the CH$_3$OH fluxes are at least a factor 10 too low. The models with 10$^{-9}$ abundance and high dust density full models underestimate all fluxes by 1 to 3 orders of magnitude. The line ratios are derived and compared with the line ratios in the RADEX plot in Figure \ref{fig:radex}. The line ratios for H$_2$CO for 10$^{-7}$ abundance are closer to the data values, suggesting that the  H$_2$CO emission is potentially optically thick (i.e. the emitting area is at least a factor 5 smaller than our estimate). The ratios shift to slightly lower values for the higher dust density models, consistent with lower temperatures. For CH$_3$OH the ratios are essentially the same for the two abundances and in similar temperature regime as the data, suggesting the emission remains optically thin for both. The ratios shift to higher values for the high dust density full models, consistent with higher temperatures, which is the result of the strong upward shift in the emitting layers. However, the column density and flux drop significantly in that case, implying that this is not a realistic scenario. Overall, the spread in temperatures for the different ratios implies that the various lines trace different temperature regions in the disk with potentially different abundances. A more detailed analysis is saved for future work.

\begin{figure*}[!ht]
\includegraphics[width=\textwidth]{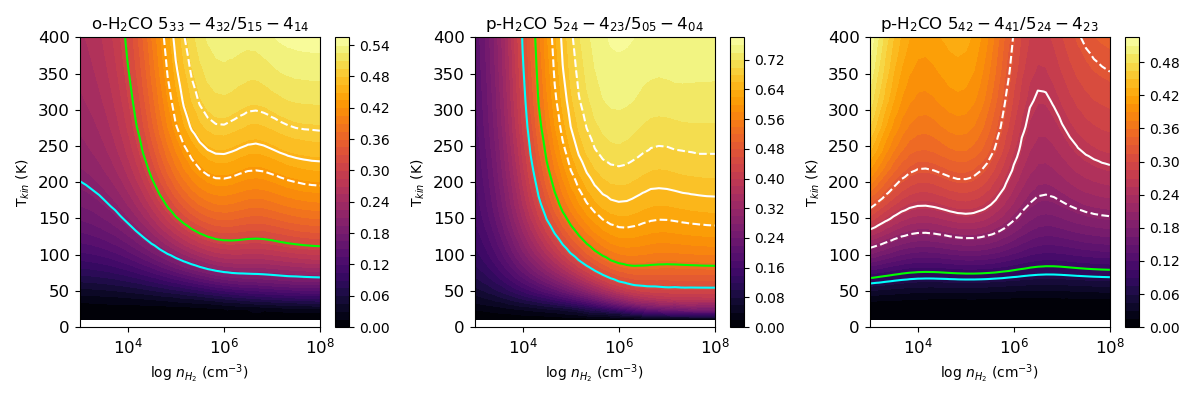}\\
\includegraphics[width=\textwidth]{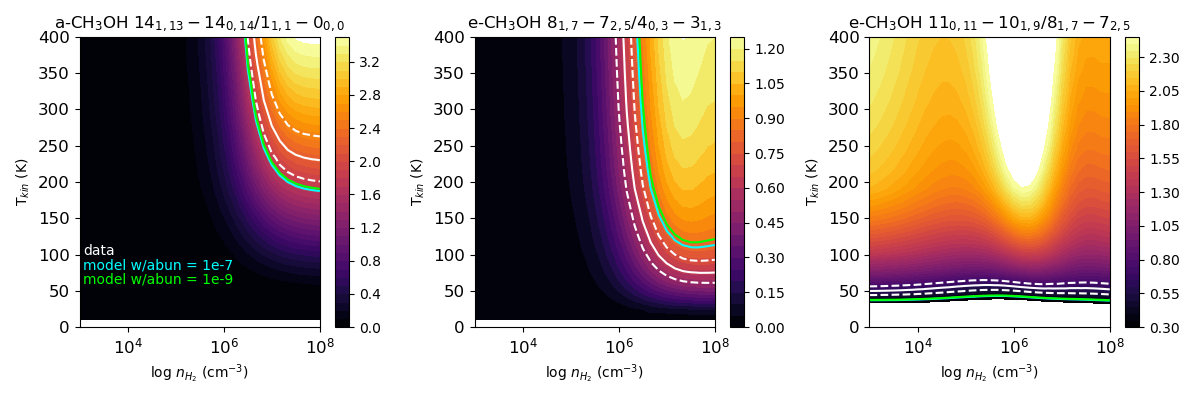}
\caption{Expected line ratios as computed by RADEX for H$_2$CO (top) and CH$_3$OH (bottom) for a column density of 10$^{14}$ cm$^{-2}$. The white contours indicate the observed values and the dashed lines the uncertainty. The coloured lines indicate the ratios as computed for our DALI models of the settled model for fixed abundances 10$^{-7}$ and 10$^{-9}$ in specific emitting layers as described in the text.}
\label{fig:radex}
\end{figure*}

\end{document}